\documentclass[11pt]{article}
\usepackage{pdfsync}
\usepackage{graphicx} 
\usepackage{subfig}
\usepackage{color}
\usepackage{amsfonts}
\usepackage{amssymb}
\usepackage{epsfig}
\usepackage{dcolumn}
\usepackage{booktabs}
\usepackage{multirow}
\usepackage{makecell}
\usepackage{boldline}
\usepackage{marvosym}

\usepackage[colorlinks]{hyperref}
\hypersetup{
    colorlinks=true,
    linkcolor=blue,
    filecolor=magenta,  
    urlcolor=[rgb]{0.00,0.00,0.50},    
    citecolor=red,
    }

\usepackage{url}

\advance \textheight by \topskip
\usepackage{amsmath}
\usepackage[english]{babel}
\makeatletter
 \@addtoreset{equation}{section}
\makeatother

\usepackage{amsmath,amssymb}
\makeatletter \@addtoreset{equation}{section}
\makeatother
\def\be{\begin{equation}}

\def\be{\begin{equation}}
\def\ee{\end{equation}}

\def\A{\mathbb A}
\def\B{\mathbb B}

\def\R{\mathbb R}
\def\W{\mathbb W}

\usepackage{amsmath,amssymb}
\def\bea{\begin{eqnarray}}
\def\eea{\end{eqnarray}}
\def\barray{\begin{array}}
\def\earray{\end{array}}

\begin{document}

\title{
{\bf 
ABC of ladder operators for
rationally extended quantum harmonic oscillator
systems 
 }}

\author{{\bf Jos\'e F. Cari\~nena${}^a$ and Mikhail S. Plyushchay${}^b$}  \\
[8pt]
{\small \textit{
${}^a$Departamento de F\'{\i}sica Te\'orica, 
Universidad de Zaragoza, 50009 Zaragoza, Spain}}\\
{\small \textit{ ${}^b$Departamento de F\'{\i}sica,
Universidad de Santiago de Chile, Casilla 307, Santiago 2,
Chile  }}\\
[4pt]
 \sl{\small{E-mails:  \textcolor{blue}{jfc@unizar.es}, 
\textcolor{blue}{mikhail.plyushchay@usach.cl}
}}
}
\date{}
\maketitle

\begin{abstract}
The problem of construction of ladder operators for
 rationally extended quantum harmonic oscillator (REQHO)
 systems of a general form is investigated in the light of
existence  of different  schemes of the
Darboux-Crum-Krein-Adler transformations
by which such systems
can be generated from the quantum harmonic oscillator. 
Any 
REQHO system is characterized by  
the number of separated states in its spectrum,
the number of `valence bands' in which the 
separated states are organized, and by the total
number of the missing energy levels
and their position.
All these peculiarities  
of a REQHO system
are shown to be detected and reflected by 
a  trinity $(\mathcal{A}^\pm$,
$\mathcal{B}^\pm$, $\mathcal{C}^\pm$)  
of the basic  (primary) lowering and raising 
ladder operators related 
between themselves by certain algebraic identities with 
coefficients   polynomially-dependent 
on the Hamiltonian.
We show that all the  
secondary, higher-order ladder operators
are obtainable  by a composition  of  
 the basic ladder operators of the trinity
which form  the set of the 
spectrum-generating operators.
Each trinity, in turn,  can be constructed from
the intertwining operators of the two 
complementary minimal schemes of the 
Darboux-Crum-Krein-Adler transformations.
\end{abstract}

\vskip.5cm\noindent

\section{Introduction}\label{secIntro}

There are two basic  exactly solvable 
quantum mechanical systems 
which reveal themselves directly or indirectly
in association with other systems
and play a fundamental role in many
physical theories and applications.
One of them is a free particle characterized
by a continuous spectrum.   The other 
one
is a harmonic oscillator with its 
infinite equidistant discrete
spectrum of the bound states.     

Free particle is essential, in particular,
for  understanding  the properties of the soliton solutions 
in  integrable systems.  Quantum reflectionless
potentials represent a snapshot of soliton solutions
to the classical KdV equation.
Initially, reflectionless potentials  
were obtained  with the help of 
the method of the inverse scattering transform in
solving the problem of theoretical
construction of a solid dielectric 
medium that is perfectly transparent to 
electromagnetic radiation \cite{KaiMos}.
This  important class of  the systems can also be 
generated from the quantum  free particle  
by means   of Darboux  transformations (DTs) and 
their generalization in the 
form   of  Darboux-Crum transformations 
(DCTs) 
\cite{Darb,Schr,InfHull,Crum,MatSal,CR00,CR01,MelRos,CR08}.
Notice here that reflectionless 
systems 
appear for instance in 
the Gross-Neveu model \cite{GroNev}
in the context of the hadron physics \cite{DasHasNev} and the 
physics of conducting polymers \cite{CamBis}.
Any quantum reflectionless system with $n$ bound states in its spectrum
is characterized by 
the presence of a nontrivial Lax-Novikov integral that is a differential operator 
of order $2n+1$ having  a structure of a Darboux-dressed
momentum operator of the free particle system.
It is this integral of motion  that distinguishes the  states in  
the doubly degenerate continuous part of the spectrum of a reflectionless system
and detects all the non-degenerate bound states 
as well as a state at the very 
edge of the continuous part of the spectrum by annihilating them
\cite{CoJaPl,LaxNov}.
It is the same  operator that plays a fundamental role 
in the theory of nonlinear integrable systems 
\cite{BBEIM,GesHol}.
 The peculiarity  of reflectionless systems 
also reveals itself in the nature of
the quantum mechanical supersymmetry associated with them.
Instead of a usual linear or non-linear $\mathcal{N}=2$ 
supersymmetric structure which appears in 
an extended system composed from a  pair
of  quantum mechanical systems related by 
a DT or DCT, the
extended  system composed from a pair of reflectionless
systems is described by the 
exotic nonlinear $\mathcal{N}=4$ supersymmetry
generated by two pairs of supercharges 
 alongside with the two bosonic integrals 
of motion constructed from  the Lax-Novikov integrals
of the subsystems \cite{LaxNov,exosusy}. The appearance of the 
exotic nonlinear supersymmetric
structure associated with reflectionless systems 
is traced  to the fact that any reflectionless Schr\"odinger system 
can be related to a free particle system  not 
only by one  but by two different DCTs due to 
the presence of the momentum operator 
in the quantum free particle system in the form of
its integral of motion~\footnote{A general  picture is more
complicated, however.  In the case of a coincidence of some or all  
discrete energy levels  of the two subsystems,  the supersymmetric 
reflectionless partners can be related directly by a DCT  of a lower order,
without a necessity to construct a chain of Darboux transformations via 
a  free particle system. This happens  due to 
the opening of a kind of a  `direct tunnelling channel' 
that can appropriately be understood from the standpoint of 
the picture of  soliton scattering, see \cite{LaxNov,Transmut}.}.

Some time ago, there  has been discovered a new very broad
class of exactly solvable systems 
which represent certain rational extensions 
of the quantum harmonic oscillator 
\cite{Dub,Adler,SamOvc,BagSam,Spir,JunRoy,CPRS,FellSmi,Sesma,GGM,Pupas}.
The eigenstates of rationally extended quantum harmonic oscillator 
(REQHO) systems are given in terms
of exceptional Hermite polynomials,
and can be obtained from  the quantum harmonic oscillator (QHO) system 
by an appropriate 
DCT, or its further generalization in the form 
of the Darboux-Crum-Krein-Adler transformation
 \cite{Adler,Krein,GomGrMil}~\footnote{Exceptional Jacobi 
and Laguerre polynomials 
\cite{Exc1,Exc1+,OdSas,OdSas+,Ques+,Ques++,SasTsZh,Grand,Sasaki}
can be associated 
with Darboux-Crum transformed free particle 
on finite interval (particle in infinite potential well) 
and isotonic oscillator, respectively; the isotonic oscillator, in turn,
 can  be related 
to the QHO by a singular Darboux transformation.}.
In what follows we shall refer to any 
generalized 
Darboux transformation 
of the QHO with
intertwining operators to be higher order differential operators
as the Darboux-Crum-Krein-Adler transformation (DCKAT).
Instead of a continuous spectrum of the free particle, 
the QHO is characterized 
by an infinite discrete spectrum of  bound states.
In spite of such a radical difference, 
the QHO  is also a very peculiar system
because  its discrete spectrum
is equidistant. As a consequence, instead of the 
Hermitian momentum operator integral
 that encodes reflectionless nature of the free particle, 
the QHO possesses a pair of
Hermitian conjugate ladder operators
which are spectrum-generating operators.  
Similarly to the relation between reflectionless 
and free particle systems,
a given REQHO system can be obtained from  
the QHO by different DCKATs. 
Then one can expect  by analogy 
with the pairs of reflectionless systems 
related by DCTs and the exotic supersymmetric structure 
associated with them that the 
REQHO systems should be characterized 
by some special properties.
Particularly, it seems to be natural to  expect 
the appearance of peculiarities 
related to the ladder operators
for such a family of quantum
systems. 

It is worth to mention  here that 
finite-gap systems, a  limit case 
of which corresponds to reflectionless systems,
and the QHO
are generated by the periodic 
Darboux chains  \cite{VesSha}. The
last construction also produces 
Painlev\'e equations \cite{VesSha,AdlPai,Hiet}, 
that are  intimately related with isomonodromic 
deformations of linear systems and integrability properties
of nonlinear systems in partial derivatives
\cite{AdlPai,Hiet}.
The REQHO systems are  
isomonodromic 
deformations of the QHO \cite{oblom}. 

Some investigations on ladder operators 
in REQHO systems have already been realized
in \cite{Spir,JunRoy,MarQue1+,MarQue1,MarQue1++,MarI,Ques,CarPly}. 
This has been done, however, for some particular examples
of the REQHO systems or for some particular families 
of such systems, while the problem of construction 
 of  ladder  operators and investigation of their properties
for REQHO systems of a general
form  remains open. 
Note also that  in the indicated 
works only some special aspects related to the ladder 
operators of the REQHO systems were studied.
 In particular, in a recent paper 
 \cite{CarPly}  we have considered 
the problem of  construction of the ladder operators 
for the simplest case of the REQHO
 by exploiting the simplest DT 
which relates the system to the QHO. 
We also investigated there the discrete chains  
related with the obtained ladder operators
given by a pair of Hermitian conjugate  
third order differential operators
   to be fermionic generators of the 
polynomially deformed
bosonized $\mathfrak{osp}(1|2)$ superalgebra.

In the present article we investigate the
problem of construction  of ladder
 operators 
for  REQHO systems 
of a general form in the light of
existence of different DCKATs by which any such a system
can be related to and generated from the QHO. 
In this point there shows up 
a similarity of the REQHO systems with 
reflectionless systems related with a free particle.
 We  
show that for any REQHO, there exists a trinity
of the basic  (primary) pairs of the   lowering and raising 
ladder operators. This trinity is proved to form the set of the 
spectrum-generating operators which  
detects and reflects all the peculiar properties of a
given REQHO  system.  

The paper is organized as follows.
In the next Section we briefly review
the properties of the DTs  
and their generalization in the form of the 
DCKATs. 
In Section \ref{secREQHOgen} we discuss 
general schemes of the DCKATs for generation 
of the REQHO systems from the QHO
by using different
sets of physical and non-physical eigenstates of  the 
latter.
Any REQHO system can be characterized by the 
total number of separated states in the low part of its 
spectrum, by the number of  the 
gapless `valence bands'
in which the eigenvalues of separated states  are organized,
and by the total number of missing energy levels 
and their position in the low part of the spectrum.  
In Section \ref{secSimplest} we consider
in detail the problem of construction of 
different  ladder operators
for the simplest REQHO system with one separated state
and two missing energy levels in a unique gap 
that separates it  from the equidistant infinite part of the spectrum.
We investigate there  the general properties
 and relations between the basic
 (primary) and  secondary, higher-order  ladder operators.  
In Section \ref{sec+2systems} we consider two more
particular examples of the REQHO systems,
one of which corresponds to a generalization of the system
from Section \ref{secSimplest}. Another example
corresponds to a REQHO system with one 
valence band composed from the two
 energy levels separated from the equidistant 
 infinite part of the spectrum by  the gap of two missing 
 energy levels. The results and observations obtained
 in Sections  \ref{secSimplest} and \ref{sec+2systems}
 are generalized then in Section \ref{sec:general}
 for the case of the REQHO systems of a general form.
 Section \ref{secSummary} is devoted to a summary of the
 obtained results,
 where we also indicate some problems 
 that could be interesting for 
further investigation.

\section{Darboux-Crum-Krein-Adler transformations}\label{secDCKATs}

Let $\psi_*(x)$ be a \emph{nodeless} physical or non-physical
\emph{real} eigenfunction of a Hamiltonian operator $H=-\frac{d^2}{dx^2}+V(x)$ 
with eigenvalue $E_*$, $H\psi_*=E_*\psi_*$.
We assume that potential $V(x)$ is a non-singular real function on all the real line
$\R$.
Define the first order differential operators
\be\label{Apsi}
A\equiv \psi_*\frac{d}{dx}\frac{1}{\psi_*}=\frac{d}{dx}-\mathcal{W}\,,
\qquad
A^\dagger=-\frac{1}{\psi_*}\frac{d}{dx}\psi_*\,,
\ee
where 
$\mathcal{W}=\frac{\psi'_*}{\psi_*}$,  
$\psi'_*=\frac{d\psi_*}{dx}$.
They   factorize the shifted Hamiltonian,
$H-E_*=A^\dagger A$,
whose potential  is given by
 $V=\mathcal{W}^2+\mathcal{W}'+E_*$ in terms of the superpotential
  $\mathcal{W}$ and factorization energy $E_*$.
The product with permuted operators  $A$ and 
$A^\dagger$ defines a supersymmetric partner
Hamiltonian, $AA^\dagger \equiv \breve{H}-E_*$,
$\breve{H}=-\frac{d^2}{dx^2}+V_*$, 
for which $\breve{V}=\mathcal{W}^2-\mathcal{W}'+E_*$.
The relation of the potential $\breve{V}(x)$ of the superpartner $\breve{H}$ 
to the potential $V(x)$ of  the system $H$ 
can be rewritten in a more convenient form
 for a further generalization,
\be\label{V*W}
\breve{V}=V-2\left(\ln \psi_*\right)''\,.
\ee
{}From factorization relations 
it follows that  the operators $A$ and 
$A^\dagger$ intertwine the partner Hamiltonians,
\be\label{A*H}
AH=\breve{H}A\,,\qquad 
A^\dagger \breve{H}=H A^\dagger\,. 
\ee
As a consequence, $A$ and 
$A^\dagger$  mutually map
the eigenstates of the superpartners.
Namely, if $\psi(x;E)$ is a physical or non-physical
eigenstate of $H$ of eigenvalue $E\neq E_*$,
$H\psi(x;E)=E\psi(x;E)$,
then $\Psi(x;E)=A\psi(x;E)$ is an eigenstate
of $\breve{H}$ of the same eigenvalue and of the same physical or 
non-physical nature.
Vice versa, if $\Psi(x;E)$ is an eigenstate
of $\breve{H}$ of eigenvalue $E\neq E_*$,
then $A^\dagger\Psi(x;E)$ is an eigenstate
of $H$ of the same eigenvalue
and of the same nature.

If $\psi(x)$ is a solution of the second order 
differential equation $H\psi(x)=E\psi(x)$ 
for arbitrary value of $E$,  a second, linearly independent 
solution of this equation is 
\be\label{deftildepsi}
\widetilde{\psi(x)}\equiv \psi(x) \int^x \frac{d\xi}{(\psi(\xi))^2}\,.
\ee
If $\psi(x)$ is a normalizable on $\R$ function, then  
$\widetilde{\psi(x)}$ is not normalizable, and vice versa. 
The properties of $A$ and $A^\dagger$ as the operators 
that mutually map the corresponding eigenstates 
$\psi(x)$ of $H$ and $\Psi(x)$ of 
$\breve{H}$  of eigenvalue $E\neq E_*$ are valid also for  
the associated  eigenstates $\widetilde{\psi(x)}$
and $\widetilde{\Psi(x)}$. 
The case $E=E_*$ in this context is different.
The eigenstate $\psi_*$ of $H$ of eigenvalue $E=E_*$ 
constitutes  the kernel of the  operator $A$,
$A\psi_*=0$. The same is valid for the 
state $\Psi_*(x)\equiv 1/\psi_*(x)$, which constitutes the 
kernel of $A^\dagger$, $A^\dagger\Psi_*=0$,
and is the eigenstate of $\breve{H}$ of the same eigenvalue $E=E_*$.
In this special case of $E=E_*$, 
the operator  $A$  transforms the state $\widetilde{\psi_*}$
into the state $\Psi_*=1/\psi_*\in\ker\,(A^\dagger)$:
\begin{equation}\label{Apsi*}
A\widetilde{\psi_*(x)}=\left(\psi_*\frac{d}{dx}\frac{1}{\psi_*}\right)\widetilde{\psi_*(x)}=
\psi_*(x)\frac{d}{dx}\int^x \frac{d\xi}{(\psi_*(\xi))^2}=
\frac{1}{\psi_*(x)}\,.
\end{equation}
Analogously, the state 
$\widetilde{\Psi_*}=\widetilde{\left(1/\psi_*\right)}$
is transformed by $A^\dagger$ into $\psi_*$ being a kernel of $A$.  
More details can be found in \cite{CarPly}
where the same notation is used.

Let us stress that for our constructions below it is important 
that 
if $\psi_*$ is a normalizable state,
then the  state $\widetilde{\psi_*}$ is not normalizable, 
and vice versa, and that the same
property  is valid  for the pair of
the states $\Psi_*=1/\psi_*$ and $
\widetilde{\Psi_*}$. 
Before  the discussion of the concrete quantum systems
we, however, completely  neglect  the questions of 
normalizability of the corresponding wave functions.
Furthermore, here and in what follows we do not 
preoccupy about normalization of the states, 
and specify wave functions   
modulo a constant multiplication factor.

The DT construction can be generalized for the case of
the DCKAT. The latter is  
generated on the basis of several seed eigenstates
$\psi_{i_1}$, $\psi_{i_2}$,\ldots,
$\psi_{i_n}$ of $H$
of  different eigenvalues 
$E_{i_k}$, $H\psi_{i_k}=E_{i_k}\psi_{i_k}$,
$k=1,\ldots, n$, with $E_{i_k}\neq E_{i_{k'}}$ for $i_k\neq i_{k'}$.
To get a nonsingular partner system $H_n$,
these states  should be  such  that 
their Wronskian $\W_n(x)\equiv \W(\psi_{i_1}(x), \psi_{i_2}(x),\ldots,
\psi_{i_n}(x))=\det\vert\vert \mathcal{F}(x)\vert\vert $,
$\mathcal{F}_{ij}=\frac{d^{i-1}}{dx^{i-1}}\psi_j$,
$i=1,\ldots, n$, $j=i_1,\ldots, i_n$,  is a nodeless function.
 The partner system $H_{n}$ is then given  by the potential
 \be\label{VnWr}
 V_n(x)=V(x)-2\left(\ln \W_n(x)\right)''.
 \ee
 The eigenstates $\psi(x;E)$ of $H$ are mapped into 
 the eigenstates $\Psi(x;E)$ of $H_n$ of the same eigenvalue $E$ 
 via the relation 
 \be\label{Wrons}
  \Psi(x;E)=\frac{\W(\psi_{i_1}(x),\ldots,\psi_{i_n}(x),\psi(x;E))}{\W_n(x)}\,.
 \ee
In the case of  $n=1$, Eq. (\ref{VnWr}) reduces to 
Eq. (\ref{V*W}) while (\ref{Wrons}) 
reduces to  the relation 
corresponding to the case of the DT  presented in the form
 $\Psi(x;E)=A_1\psi(x;E)$,
with $\psi_{i_1}=\psi_*$, $A_1=A$ and $H_1=\breve{H}$.
 Furthermore, 
for $n>1$
Eq. (\ref{Wrons}) can 
be presented in a form that generalizes 
the indicated DT's formula. 
For this  we iteratively define
 two sequences of related 
differential operators $A_m$, $m=1,\ldots$,
and 
$\A_m$, $m=0,1,\ldots,$
 as follows: $\A_0\equiv 1$,
\be\label{Amdef}
A_m=\left(\A_{m-1}\psi_{i_m}\right)\frac{d}{dx}\frac{1}{\left(\A_{m-1}\psi_{i_m}\right)}\,,
\qquad
m=1,2,\ldots,
\ee
\be\label{defAn}
\A_n=A_n A_{n-1}\ldots A_1\,.
\ee 
Let us denote  $H_0\equiv H$
and define  $H_{m-1}=A_m^\dagger A_m+E_{i_m}$.
We have  
 $A_m A_m^\dagger=H_{m}-E_{i_m}$,
 and  obtain a generalization of 
the intertwining relations (\ref{A*H}),
$A_m H_{m-1}=H_m A_m$,
$A_m^\dagger H_{m}=H_{m-1} A_m^\dagger$.
Then  we can present 
  (\ref{Wrons}) 
in the equivalent form
\be\label{psiAnpsi}
\Psi(x;E)=\A_n\psi(x;E)\,. 
\ee

The $n$-th order differential 
operators $\A_n$ and $\A_n^\dagger$ 
intertwine the partner systems $H_0$ and $H_n$,
\be\label{AnH0Hn}
\A_n H_0=H_n\A_n\,,\qquad
\A_n^\dagger H_n=H_0\A_n^\dagger\,.
\ee
The products of  operators   $\A_n$ and $\A_n^\dagger$ 
turn out to be polynomials in the corresponding Hamiltonian operators
with  roots equal  to the energies of factorization:
\be
\A_n^\dagger \A_n=\prod_{k=1}^n (H_0-E_{i_k})\,,\qquad
\A_n \A_n^\dagger =\prod_{k=1}^n (H_n-E_{i_k})\,.
\ee

For non-singular partners
$H_0$ and $H_n$, some or all of the 
intermediate second-order differential operators 
$H_m$ with $m=1,\ldots,n-1$
can be singular.
If we change the order of the 
seed states, the Wronskian is left invariant 
modulo a possible multiplication by $(-1)$,
that does not change Eqs. (\ref{VnWr}) and
(\ref{Wrons}). Differential operator $\A_n$ 
 is not changed either, and  from (\ref{Wrons}) 
 one can conclude   that  
the kernel of $\A_n$ is spanned by 
 the complete set of the seed states, 
 $\ker\, (\A_n)=\text{span}\, \{\psi_{i_1},\ldots,\psi_{i_n}\}$. 
However, under  permutation in the order
of the seed eigenstates 
the first order differential operators
entering into the factorized form (\ref{defAn}) 
of $\A_n$  are  changed. 
As a result, the nature of some or of all
of the intermediate
systems $H_m$ can be changed from 
a non-singular (singular) to a singular (non-singular).

\section{Generation of  the REQHO systems from the QHO}\label{secREQHOgen}

Let us turn now to the  specific example of the QHO
system 
given by the  Hamiltonian operator 
$
H_{\textrm{osc}}=-\frac{d^2}{dx^2}+x^2.
$
Its bound  eigenstates are described 
by (not normalized here)  wave functions 
 \be\label{psinHosc}
\psi_n(x)=H_n(x)e^{-x^2/2}
\ee
which correspond to eigenvalues
$E_n=2n+1$, $n=0,1,\ldots$, where $H_n(x)$ are 
Hermite polynomials. 
The change of variable $x\rightarrow i\,x$
 generates a change
of the sign of the Hamiltonian,
$H_{\textrm{osc}}\rightarrow -H_{\textrm{osc}}$,  and so,  transforms physical eigenstates $\psi_n(x)$ 
 into non-physical eigenstates  
$
 \psi^-_n(x)=\mathcal{H}_n(x)e^{x^2/2}
 $
 of $H_{\textrm{osc}}$ of eigenvalues $E_n^-=-(2n+1)$, $n=0,1,\ldots$, where 
 $\mathcal{H}_n(x)=H_n(i\,x)$.
Unlike $H_n(x)$,  the polynomials $\mathcal{H}_n(x)$ with even index have 
no real zeros
 while the unique real zero of these polynomials with odd index is 
 at $x=0$.  More details can be found in \cite{CarPly}.
 
The choice  of the ground state
$\psi_0=e^{-x^2/2}$ 
 as a seed eigenfunction $\psi_*$ for the DT in (\ref{Apsi})
generates the first order differential operators
\be
A_1=\frac{d}{dx}+x\equiv a^-\,,\qquad
A^\dagger_1=-\frac{d}{dx}+x\equiv a^+\,.
\ee
They factorize the shifted Hamiltonian, 
$a^+a^-=H_{\textrm{osc}}-1$, and satisfy the commutation relation
\be
[a^-,a^+]=2\,.
\ee
As a consequence  we have 
\be\label{Hosca+-}
[H_{\textrm{osc}},a^\pm]=\pm 2 a^\pm\,.
\ee
The two relations in (\ref{Hosca+-})  can be rewritten equivalently in the form
\be\label{H0a+-}
a^-H_{\textrm{osc}}=(H_{\textrm{osc}}+2)a^-\,,\qquad
a^+(H_{\textrm{osc}}+2)=H_{\textrm{osc}} a^+ \,,
\ee
and mean that $a^-$ and $a^+$ are  the ladder lowering and 
raising operators of the QHO.
Relations in (\ref{H0a+-}) can also be interpreted
as that $a^-$ and $a^+$ intertwine 
the QHO system $H_{\textrm{osc}}=H_0$ with the SUSY-partner system 
$H_1=H_{\textrm{osc}}+2$
which is just the shifted
QHO.
Since $a^-\psi_0=0$, 
from the point of view of the DT
one can consider the shifted system
$H_1$ as the QHO with the removed ground 
state. But since $a^-\psi_n=\psi_{n-1}$,
the partner system $H_1$
is  the same QHO 
with the Hamiltonian shifted for $+2$. 
{}In this DT picture, the ground state $\psi_0$ of $H_{\textrm{osc}}$
is created by application of $A_1^\dagger=a^+$
to the function  
$\widetilde{\psi^-_0}=\widetilde{\left(1/\psi_0\right)}$
that  is  a non-physical eigenstate of $H_1$ of 
eigenvalue $E=1$.

The DCKAT generated on 
the basis of the set of the seed eigenstates 
$\psi_0,\ldots,\psi_{n-1},\psi_n$,
$n=1,\ldots,$
produces the intertwining operators
$\A_{n+1}^-=(a^-)^{n+1}$ and 
$\A^+_{n+1}=\A_{n+1}^\dagger=(a^+)^{n+1}$,
which intertwine the QHO with the 
SUSY-partner system
$H_{n+1}=H_{\textrm{osc}}+2(n+1)$,
\be
\A_{n+1}^-H_{\textrm{osc}}=H_{n+1}\A_{n+1}^-,\qquad
\A_{n+1}^+ H_{n+1}=H_{\textrm{osc}}\A_{n+1}^+\,.
\ee
The case $n=0$ here reproduces  
the relations of the DT generated by
the choice  $\psi_*=\psi_0$.
Note that in the case of $n=1$ 
the permutation of the seed states
in the DCKAT construction,
$(\psi_0,\psi_1)\rightarrow (\psi_1,\psi_0)$,
gives rise to a singular 
operator 
\be\label{A1iso}
A_1=\psi_1\frac{d}{dx}\frac{1}{\psi_1}=
\frac{d}{dx}+x-\frac{1}{x}\equiv a^-_{\textrm{iso}}\,.
\ee
 This operator 
acting on the second 
chosen seed eigenstate
$\psi_0$ gives a function $-\frac{1}{x}e^{-x^2/2}=-1/\psi_1^-$, 
which according to 
(\ref{Amdef}) generates the second 
factorization operator
$A_2=\frac{d}{dx}+x+\frac{1}{x}$ that also is singular.
The product of these two singular operators
gives the same second-order non-singular intertwining 
operator~\footnote{For the discussion 
of related phenomena in a context of the quantum 
second-order supersymmetry 
anomaly and coupling-constant metamorphosis see ref. \cite{SchwAn}}
 as  the scheme
with the $(\psi_0,\psi_1)$ pair:
$A_2A_1=(a^-)^2=\A_2^-$. 
The intermediate Hamiltonian 
$H_1$ in this case is a singular at $x=0$ operator
corresponding to the (shifted)  quantum isotonic oscillator,
$A_1^\dagger A_1=H_{\textrm{osc}}-3$,
$A_1 A_1^\dagger=H_1-3$,
which is given by the  potential 
$V_1(x)=x^2+\frac{2}{x^2}+2$.
We also have here the relations 
$A_2^\dagger A_2=H_1-1$,
$A_2A^\dagger_2=H_{\textrm{osc}}+3$.

The choice of  a nodeless non-physical state $\psi_*=\psi^-_0$ 
corresponding to factorization energy $E_*=-1$
in the DT construction gives 
\be
A_1=\frac{d}{dx}-x=-a^+\,,\qquad
A_1^\dagger=-a^-\,.
\ee
We obtain the same ladder operators of the QHO,
but now they will intertwine the QHO with the
SUSY-partner system $H_1=H_{\textrm{osc}}-2$, 
\be\label{H0a+-*}
a^+H_{\textrm{osc}}=(H_{\textrm{osc}}-2)a^+\,,\qquad
a^-(H_{\textrm{osc}}-2)=H_{\textrm{osc}} a^- \,.
\ee
Here the action of $A_1$ on the non-physical eigenstate
$\widetilde{\psi^-_0}=\widetilde{\left(1/\psi_0\right)}$
 of $H_0=H_{\textrm{osc}}$ of 
eigenvalue $E=-1$ produces a physical ground state $\psi_0$
for the partner system  $H_1$. 

In the case of the DCKAT generated
on the basis of the seed eigenstates
$\psi_0^-,\ldots,\psi_{n-1}^-,\psi_n^-$,
$n=1,\ldots$,
we obtain a partner system with the added 
$n+1$ bound states in the lower part of the spectrum
of the QHO which is described by the shifted
QHO Hamiltonian $H_{n+1}=H_{\textrm{osc}}-2(n+1)$. 

The described 
DCKATs based on the choice of the seed eigenstates
$\psi_0,\ldots,\psi_n$ or  $\psi^-_0,\ldots,\psi^-_n$
reflect the property of the special shape invariance of the 
QHO system.

\vskip0.1cm

Before we proceed further, let us summarize  
briefly
the main features of the non-singular 
DT  and DCKAT schemes
based on other choices of the sets of 
physical, $\psi_n$, and non-physical, $\psi^-_n$,
eigenstates of the QHO as the seed eigenstates
\cite{GGM,MarQue1,Ques}.
This will generalize  the preceding discussion and will
allow us  to generate the REQHO systems. 

\vskip0.1cm

Consider the DCKAT generated on the basis of  the 
physical eigenstates $\psi_{i_1}$,
$\psi_{i_2}$, $\ldots$, $\psi_{i_n}$, and non-physical 
eigenstates $\psi_{j_1}^-$,
$\psi_{j_2}^-$, $\ldots$, $\psi_{j_l}^-$,
 of the QHO.
The peculiarity of the physical and non-physical eigenstates in both
 families is that they have  
a  form of polynomials  multiplied by exponential 
functions $e^{-x^2/2}$ and $e^{x^2/2}$, respectively.
As a result we obtain a quantum system described
by a potential to be a rational function.
In order 
a new quantum system generated by the DCKAT be non-singular, 
the complete set of the seed  states has to be composed from the 
blocks of the states
$(\psi_0^-,\ldots,\psi_{j_1-1}^-,\psi_{j_1}^-)$, $(\psi_{j_2}^-,\ldots,
\psi_{j_2+l_2}^-),\ldots,$
$(\psi_{j_r}^-,\ldots,
\psi_{j_r+l_r}^-)$,
$(\psi_0,\ldots,\psi_{i_1-1},\psi_{i_1})$,
$(\psi_{i_2},\ldots,
\psi_{i_2+2m^+_2+1}),\ldots,$
$(\psi_{i_s},\ldots,
\psi_{i_s+2m^+_s+1})$,
where $j_2=j_1+2m^-_1+3$, $j_{k+1}= j_{k}+2m^-_k+3$,
$i_1< i_2$, $i_k+2m^+_k+1<i_{k+1}$,
and $m^+_k$, $m^-_k$ and $l_k$ can take values $0,1,\ldots$.
The total number of such blocks can be arbitrary and 
blocks that include the states $\psi_0$ and $\psi^-_0$ 
can be absent.
Up to a possible  constant shift, 
the  generated system will have the gapped spectrum 
of the QHO with the deleted levels  appearing
at the positions of energies 
 corresponding to  physical states $\psi_i$
in these blocks, and with new, added energy levels 
appearing at the positions of
energies of non-physical eigenstates $\psi^-_j$.
In other words, the inclusion of physical states $\psi_i$ into the 
generating set of the seed states eliminates the energy levels,
while the inclusion of non-physical states $\psi^-_j$ introduces 
corresponding additional energy levels into the spectrum.
Each gap in the spectrum of the resulting system 
contains an even number of the missing energy levels. 
As in the simplest examples we considered above
with the partner system to be the same but the shifted 
QHO, the same REQHO system
can be produced by DCKATs based on different
choices of the sets of the seed eigenstates. 
Different choices of the seed states 
generate different  intertwining operators
which relate REQHO system with the 
QHO. As a consequence, as we shall see, 
 there exist different 
ladder operators for the 
same REQHO, which possess  different 
properties. 
Below we first consider some simple concrete examples
of the REQHO systems. This will allow us
to investigate in detail the families
of the DCKATs associated with a given REQHO system, to 
identify different ladder operators, and 
to study their properties as well as  
to establish the relations between them.
Then the results will be developed 
for the  case of the REQHO systems of a general form.

\section{Simplest REQHO system and its ladder operators}\label{secSimplest}

A  simplest REQHO system 
can be produced by
taking a nodeless non-physical eigenstate of 
energy $E=-5$ of $H_{\textrm{osc}}$, 
\be
\psi^-_2=(1+2x^2)e^{x^2/2}\,.
\ee 
The first order differential   operators 
\be\label{A+-R}
A^-\equiv\psi_2^-\frac{d}{dx}\frac{1}{\psi^-_2}=\frac{d}{dx}-x-\frac{4x}{2x^2+1}\,,
\qquad A^+=(A^-)^\dagger\,
\ee
factorize the shifted QHO Hamiltonian 
\be\label{A+A-defH}
A^+A^-=-\frac{d^2}{dx^2}+x^2+5=H_{\textrm{osc}}+5\equiv H\,.
\ee
Their permuted product  generates a simplest REQHO system, 
\be\label{breveHdef} 
A^-A^+=-\frac{d^2}{dx^2}+x^2+3 +8\frac{2x^2-1}{(2x^2+1)^2}\equiv \breve{H}\,.
\ee
We have the  intertwining relations
\be\label{AHinter}
A^-H=\breve{H}A^-\,,\qquad
A^+\breve{H}=H A^+\,,
\ee
from which it follows that the systems $H$ and $\breve{H}$ are almost 
isospectral, and  operators (\ref{A+-R}) 
provide a  map  between the eigenstates 
of the QHO and the REQHO systems.
The excited eigenstates 
of $\breve{H}$ are the bound states 
\be\label{PsinA}
\Psi_n(x)=A^-\psi_n(x)\,,\qquad
E_n=6+2(n-1)\,,\qquad
n=1,2,\ldots,
\ee
where $\psi_n(x)$ are the QHO eigenstates (\ref {psinHosc}).
In correspondence with the general relation (\ref{Apsi*}),
the ground state and its energy are
\be\label{APsi0}
\Psi_0=A^-\widetilde{\psi^-_2}=\frac{1}{\psi^-_2}\,,\qquad
E_0=0\,.
\ee
In correspondence with
a general picture described above, this state constitutes the 
kernel of the operator $A^+$.
The ladder operators for $\breve{H}$ can be constructed 
by the Darboux-dressing 
of the ladder operators $a^\pm$ of the QHO,
\be\label{ladderA1}
\mathcal{A}^\pm=A^-a^\pm A^+\,.
\ee
We have 
\be\label{HcomA}
[\breve{H},\mathcal{A}^\pm]=\pm 2\mathcal{A}^\pm\,,
\ee
and
\be\label{A+A-Hbr}
\mathcal{A}^+\mathcal{A}^-=\breve{H}(\breve{H}-2)(\breve{H}-6)\,,\qquad
\mathcal{A}^-\mathcal{A}^+=\breve{H}(\breve{H}+2)(\breve{H}-4).
\ee
Due to the last factor in (\ref{ladderA1}) and in correspondence 
with relations (\ref{A+A-Hbr}), 
the ground-state of
the REQHO of zero energy, $\Psi_0=1/\psi_2^-$,
 is annihilated 
by both ladder operators $\mathcal{A}^-$ and $\mathcal{A}^+$.
Another peculiarity is 
that the kernel of the lowering operator 
$\mathcal{A}^-$ also contains 
the first excited physical state
$\Psi_1=A^-\psi_0$ of energy $E=6$,
and the non-physical state
$A^-\psi^-_1$ which is the eigenstate of $\breve{H}$ 
of the eigenvalue $E=2$.
The three-dimensional  kernel
of the  lowering ladder operator $\mathcal{A}^-$
is therefore
\be\label{kernelA-}
\ker\,(\mathcal{A}^-)=\text{span}\,\{\Psi_0,A^-\psi^-_1,
\Psi_1\}\,.
\ee
Besides the 
ground state $\Psi_0$,
the kernel of $\mathcal{A}^+$   includes 
the  two
states $A^-\psi^-_3$ and $A^-\psi^-_0$,
which are non-physical  eigenstates of $\breve{H}$ 
of the eigenvalues $-2$ and $4$,
\be\label{kernelA+}
\ker\,(\mathcal{A}^+)=\text{span}\,\{\Psi_0,A^-\psi^-_3,
A^-\psi_0^-\}\,.
\ee

We denote $(\alpha_1)$ this scheme based on the DT with
generating function $\psi^-_2$, 
\be
(\alpha_1)=\{\psi^-_2\}.
\ee
Up to a global shift,  the same REQHO system can also 
be produced
by means of any of the DCKAT  schemes
\be
(\alpha_2)=\{\psi^-_0,\psi^-_3\}\,, \ldots, 
(\alpha_{n+1})=\{\psi^-_0,\psi^-_1,\ldots,
\psi^-_{n-1},\psi^-_{n+2}\}\,.
\ee
The presence of the first $n$ states 
$\psi^-_0,\psi^-_1,\ldots,
\psi^-_{n-1}$
in the scheme
$(\alpha_{n+1})$ gives rise to 
the addition  of the corresponding energy levels 
into the spectrum of the QHO while the inclusion 
of the state $\psi^-_{n+2}$ results finally in the generation  
of the shifted gapped REQHO system $H_{n+1}=\breve{H}-2n$.
The intertwining operators between 
$H$ defined by (\ref{A+A-defH}) and $H_{n+1}$  in this case are 
\be\label{Aalpha_n} 
\A_{n+1}=A^-(a^+)^n\,,\qquad
 \A_{n+1}^\dagger=(a^-)^nA^+\,,
 \ee 
 where  $A^-$ and $A^+$ are 
 given by Eq. (\ref{A+-R}).
 Combining the intertwining operators of the schemes
 $(\alpha_{n+1})$ and $(\alpha_1)$,
 we can construct the 
 higher-order  ladder operators
 \be\label{laddern+1}
  A^-\A_{n+1}^\dagger =A^-(a^-)^nA^+\equiv \mathcal{A}_n^-\,,\qquad
 \A_{n+1}A^+=A^-(a^+)^nA^+=(\mathcal{A}_n^-)^\dagger\equiv \mathcal{A}_n^+\,,
 \ee
 $[\breve{H}, \mathcal{A}_n^\pm]=\pm 2n \mathcal{A}_n^\pm$,
 where $\mathcal{A}^\pm_1=\mathcal{A}^\pm$.
 This in particular
  means  that the third-order differential  operators
 (\ref{ladderA1}), which have the nature of the 
 Darboux-dressed QHO operators $a^\pm$, 
 can also be considered as the ladder operators
 generated via a composition of the intertwining 
 operators corresponding to the schemes 
 $(\alpha_1)$ and $(\alpha_2)$,
   $\mathcal{A}^-=A^-(a^-A^+)$,
 $\mathcal{A}^+=(A^-a^+)A^+$.
 The use of the scheme $(\alpha_n)$ with $n>2$ 
 instead of $(\alpha_2)$ in such a composition 
 provides  us therefore with the  
 $(\alpha_1)$-Darboux-dressed form 
 (\ref{laddern+1})
 of the higher-order
 ladder operators $(a^\pm)^n$ of the QHO.
The following relations 
between $\mathcal{A}^-$ and $\mathcal{A}^-_n$  are valid, 
 \be\label{A-n=An-} 
 (\mathcal{A}^-)^n=\prod_{j=1}^{n-1}(\breve{H}+2j)\cdot \mathcal{A}^-_n=
  \mathcal{A}^-_n\cdot \prod_{j=1}^{n-1}(\breve{H}-2j)\,,
 \ee
and analogous relations are obtained from them 
for $\mathcal{A}^+$ and $ \mathcal{A}^+_n$ 
by  the Hermitian conjugation.
In 
a more general case the composition of the intertwining operators of the 
schemes 
$(\alpha_n)$ and $(\alpha_m)$ with $n>m$ generates 
the higher-order ladder operators 
$\mathcal{A}^\pm_{n-m}$,
\be
\A_{m+1}\A_{n+1}^\dagger=\prod_{j=1}^m(\breve{H}-4-2j)\cdot \mathcal{A}_{n-m}^-\,,
\qquad
\A_{n+1}\A_{m+1}^\dagger= \mathcal{A}_{n-m}^+\cdot \prod_{j=1}^m(\breve{H}-4-2j)\,.
\ee

The REQHO system (\ref{breveHdef} ) 
can also be generated 
via the DCKAT based on  the 
physical eigenstates $\psi_1$ and $\psi_2$.
We denote this scheme, 
which eliminates two neighbour energy levels
$E=8$ and $E=10$ 
in  the spectrum of the shifted 
QHO (\ref{A+A-defH}),
 as $(\beta_2)$:
\be\label{beta2}
(\beta_2)=\{\psi_1,\psi_2\}\,.
\ee
We denote  $\B^\pm_2$ the second-order
 intertwining operators $\A_2$ and $\A_2^\dagger$
constructed 
on the basis of these two states according to (\ref{defAn}), 
\be
\B_2^-\equiv A^-_{\textrm{iso}}a^-_{\textrm{iso}}\,,\qquad
\B_2^+\equiv (\B_2^-)^\dagger=a^+_{\textrm{iso}}A^+_{\textrm{iso}}\,.
\ee
The operator $a^-_{\textrm{iso}}$ is defined in Eq. (\ref{A1iso}),
and $a^+_{\textrm{iso}}=(a^-_{\textrm{iso}})^\dagger$.
The result of the action of the 
operator $a^-_{\textrm{iso}}$ on the
QHO's eigenstate $\psi_2$ can be presented 
in terms of its physical and non-physical  eigenstates in the 
form
$a^-_{\textrm{iso}}\psi_2=-\psi_0\psi^-_2/\psi^-_1\equiv \phi$.
The 
first-order differential operators 
$A^\pm_{\textrm{iso}}$  are  generated by the function $\phi(x)$,  
according to  (\ref{Amdef}),
\be
A^-_{\textrm{iso}}=\phi(x)\frac{d}{dx}\frac{1}{\phi(x)}=\frac{d}{dx}
+x+\frac{1}{x}-\frac{4x}{1+2x^2}\,,\qquad
A^+_{\textrm{iso}}=(A^-_{\textrm{iso}})^\dagger\,.
\ee
We have 
\be\label{aisodef}
a^+_{\textrm{iso}}a^-_{\textrm{iso}}=H-8\,,\qquad
a^-_{\textrm{iso}}a^+_{\textrm{iso}}=H_{\textrm{iso}}\,,
\ee
where 
\be\label{Hisodef}
H_{\textrm{iso}}=-\frac{d^2}{dx^2}+x^2+\frac{2}{x^2}-1
\ee
is the shifted isotonic oscillator 
to be singular at $x=0$,
and $H$ corresponds to the shifted
Hamiltonian of the QHO defined in (\ref{A+A-defH}).
We also have the relations
\be\label{Aisodef}
A^+_{\textrm{iso}}A^-_{\textrm{iso}}=H_{\textrm{iso}}-2\,,\qquad
A^-_{\textrm{iso}}A^+_{\textrm{iso}}=\breve{H}-4\,,
\ee
where $\breve{H}$ is the Hamiltonian 
of the REQHO system defined in Eq. (\ref{breveHdef}).
{}From (\ref{aisodef}) and (\ref{Aisodef})
we find
\be\label{B2+6}
\B_2^- H=(\breve{H}+6)\B_2^-\,,\qquad
\B_2^+ (\breve{H}+6)=H\B_2^+\,,
\ee
and 
\be
\B_2^+ \B_2^-=(H-8)(H-10)\,,\qquad
\B_2^-\B_2^+=(\breve{H}-2)(\breve{H}-4)\,.
\ee
By the construction, $\ker\,(\B_2^-)=\text{span}\,\{\psi_1,\psi_2\}$. 
One can also see that 
$\ker\,(\B_2^+)=\text{span}\,\{A^-\psi^-_0,A^-\psi^-_1\}=
\text{span}\,\{\B^-_2\widetilde{\psi_1},
\B^-_2\widetilde{\psi_2}\}=
\text{span}\,\{\frac{\psi_1}{\W_2},\frac{\psi_2}{\W_2}\}$,
where $\W_2(x)=\W(\psi_1,\psi_2)(x)=-\psi^-_2(x)\, e^{-\frac{3}{2}x^2}$.
\vskip0.1cm

The DCKATs corresponding to the 
$(\alpha_1)$- and $(\beta_2)$-schemes  are
in some sense complementary. The $(\alpha_1)$-scheme introduces   
effectively a new energy level into the 
spectrum of the QHO  below its ground-state
energy at the  distance equal to the tripled
distance between equidistant energy levels.
The $(\beta_2)$-scheme 
makes a similar job but by deleting the 
first two excited energy levels in the spectrum 
of the QHO. Since the Wronskian 
in the DCKAT in the 
latter scheme includes the additional 
exponential factor $e^{-\frac{3}{2}x^2}$ in comparison
with the structure of the non-physical eigenstate
$\psi^-_2$,  this 
produces the additional constant shift $+6$ 
in the potential generated by means of 
relation (\ref{VnWr})
that is reflected in  Eq. (\ref{B2+6}),
cf. Eq. (\ref{AnH0Hn}).
{}From Eq. (\ref{B2+6}) it follows that if $\psi(x;E)$
is an eigenstate of $H$ of energy $E$, 
then $\B^-_2\psi(x;E)$ is an 
eigenstate of $\breve{H}$ of energy $(E-6)$.
As a consequence,   
all the spectrum of the system generated  by 
the $(\beta_2)$-scheme will be shifted 
for  $-6$ in comparison 
with the spectrum of the 
REHQO system (\ref{breveHdef}) produced via
the $(\alpha_1)$-scheme.
In correspondence with this picture, 
the ground-state 
$\Psi_0$ of $\breve{H}$ can alternatively
be constructed from the QHO
ground-state $\psi_0$, $\Psi_0=\B^-_2\psi_0$,
cf. (\ref{APsi0}).
The excited states $\Psi_{n+1}$ of
energy $E_{n+1}=6+6n$ with $n=0,\ldots$ 
can be presented in the alternative to  (\ref{PsinA}) form
 $\Psi_{n+1}=\B^-_2\psi_{n+3}$.
All this gives a possibility 
for the construction of another pair of  ladder
operators for the REQHO system (\ref{breveHdef})
with the properties rather different to
 those of the ladder operators we obtained
by using only the $(\alpha)$-schemes. For this we take now
the composition of the intertwining operators
of the $(\alpha_1)$- and $(\beta_2)$- schemes
to construct the operators
\be\label{CB2A}
\mathcal{C}^-=\B_2^-A^+\,,\qquad
\mathcal{C}^+=A^-\B^+_2\,.
\ee
Instead of (\ref{HcomA}) and (\ref{A+A-Hbr}),
they satisfy the relations 
\be\label{HC+-3}
[\breve{H},\mathcal{C}^\pm]=\pm 6\,\mathcal{C}^\pm\,,
\ee
and 
\be\label{B+B-Hbr}
\mathcal{C}^+\mathcal{C}^-=\breve{H}(\breve{H}-8)(\breve{H}-10)\,,\qquad
\mathcal{C}^-\mathcal{C}^+=(\breve{H}+6)(\breve{H}-2)(\breve{H}-4)\,.
\ee
Like $\mathcal{A}^\pm$, these are  third-order
 differential
operators of the nature of  ladder operators.
However, acting on eigenstates of the REQHO system 
$\breve{H}$, they change the energies not in $2$ but in $6$.
In this aspect they are somewhat  similar to the 
higher-order ladder operators $\mathcal{A}^\pm_3$,
which are
fifth-order differential operators discussed
above. The essential
difference of $\mathcal{C}^-$ 
from $\mathcal{A}^-_n$ and in particular $\mathcal{A}^-$
 is that 
in correspondence with the
first relation from (\ref{B+B-Hbr}),
the kernel of $\mathcal{C}^-$ 
is composed only from physical eigenstates
of $\breve{H}$,  
\be
\ker\,(\mathcal{C}^-)=\text{span}\,\{\Psi_0,\Psi_2,\Psi_3\}\,.
\ee
The energies $0$, $8$ and  $10$ of these states are the roots
of the third-degree polynomial in the first identity in (\ref{B+B-Hbr}). 
Also, unlike $\mathcal{A}^+$,
 the kernel of the raising operator $\mathcal{C}^+$
is composed only from the
non-physical eigenstates of $\breve{H}$, 
\be\label{kerB+}
\ker\,(\mathcal{C}^+)=\text{span}\,\{A^-\psi^-_5,A^-\psi^-_1,A^-\psi^-_0\}=
\text{span}\,\{\B^-_2\psi^-_3,\B^-_2\widetilde{\psi_1},\B^-_2\widetilde{\psi_2}\}.
\ee
In correspondence with the second 
relation in (\ref{B+B-Hbr}),
the eigenvalues of the states in (\ref{kerB+}) are
$-6$, $2$ and  $4$.
{}From the point of view of the structure
of the kernels and commutation relations
(\ref{HC+-3}), 
 the ladder operators
$\mathcal{C}^\pm$ are 
similar to the 
third-order  differential 
operators $(a^\pm)^3$ in the QHO system.
However, unlike $\mathcal{C}^-$,  the operator  
$(a^-)^3$  annihilates the three lowest physical 
eigenstates of 
the QHO of the three subsequent  values of energy. 
The first exited state $\Psi_1$ of the REQHO 
system of energy $E=6$ does not belong  to 
the kernel of $\mathcal{C}^-$ and is annihilated by 
$(\mathcal{C}^-)^2$: 
$\mathcal{C}^-\Psi_1=\Psi_0$, $(\mathcal{C}^-)^2\Psi_1=0$.

The following relations can be 
established by comparing  
the kernels of the operators on both sides of
the  equalities,
\be\label{ABa3}
A^+\B^-_2=-(a^-)^3\,,\qquad
\B^+_2A^-=-(a^+)^3\,.
\ee
These  and their analogous  relations  for
other REQHO systems will play important role
in what follows. {}From them one can find 
in particular the operator identities 
$a^-\B^+_2=-(a^+)^2A^+$,
$(a^-)^2\B^+_2=-a^+A^+(\breve{H}-2)$,
as well as the Hermitian conjugate ones.

 One can introduce  additionally the operators
 $a^\pm$ inside the factorized 
 structure of the  operators $\mathcal{C}^\pm$.
 In this way one can construct the operators
 \be\label{Bn+1def}
 \mathcal{C}^-_{n+1}=\B_2^-(a^-)^nA^+\,,\qquad
  \mathcal{C}^+_{n+1}=A^-(a^+)^{n}\B^+_2\,,\quad
  n=0,\ldots\,, 
  \ee
  with the implied identification  $\mathcal{C}^\pm_1=\mathcal{C}^\pm$
  for $n=0$. 
  They satisfy the relation $[\breve{H},\mathcal{C}^\pm_{n+1}]=
  \pm (6+2n)\mathcal{C}^\pm_{n+1}$.
  These operators can be treated either as 
  the QHO operators $(a^\pm)^n$ dressed
  by the intertwining generators 
  of the $(\alpha_1)$ and $(\beta_2)$ schemes,
  or as the operators produced by
  intertwining operators (\ref{Aalpha_n}) from the
  $(\alpha_{n+1})$ scheme and those
  from the same $(\beta_2)$ scheme. 
  The kernel of $\mathcal{C}^-_{n+1}$ is 
 composed only by
the physical eigenstates of $\breve{H}$, while
the kernel of   $\mathcal{C}^+_{n+1}$ is 
spanned only by its 
non-physical eigenstates.
For instance, $\ker\,(\mathcal{C}^-_2)=\text{span}\,\{
\Psi_0,\Psi_1,\Psi_3,\Psi_4\}$.
With the help of identities (\ref{ABa3}) we also find that
$(\mathcal{C}^\pm)^n=(-1)^{n+1}\mathcal{C}^\pm_{3(n-1)+1}$, 
$n=1,\ldots\,.$

Analogously to $\mathcal{A}^\pm$, we also
introduce   the operators
\be\label{B2aB2}
\mathcal{B}^\pm=\B^-_2 a^\pm\B^+_2\,.
\ee
Unlike the third-order ladder operators 
$\mathcal{A}^\pm$ and $\mathcal{C}^\pm$, 
the $\mathcal{B}^\pm$
are fifth-order differential operators. 
They satisfy the relations 
\be
[\breve{H},\mathcal{B}^\pm]=\pm 2\mathcal{B}^\pm
\ee
and 
\be\label{B+newB-}
\mathcal{B}^+\mathcal{B}^-=
\breve{H}(\breve{H}-2)(\breve{H}-6)(\breve{H}-4)^2\,,\qquad
\mathcal{B}^-\mathcal{B}^+=
\breve{H}(\breve{H}+2)(\breve{H}-4)(\breve{H}-2)^2\,.
\ee
The kernel of $\mathcal{B}^-$ involves
two  physical 
and three non-physical eigenstates 
of $\breve{H}$,
\be\label{Bnew-ker}
\ker\,(\mathcal{B}^-)=\text{span}\,\{
\Psi_0,\Psi_1,A^-\psi^-_0,A^-\psi^-_1,A^-\widetilde{\psi^-_0}\}\,.
\ee
The eigenvalue $E=4$ of the 
non-physical eigenstates 
$A^-\psi^-_0$ and $A^-\widetilde{\psi^-_0}$ 
in the kernel of 
$\mathcal{B}^-$ 
corresponds to the
double root of the last factor in the first relation in  
(\ref{B+newB-}). 
 The kernel of the increasing ladder operator
 $\mathcal{B}^+$
 includes only one physical eigenstate,
\be\label{Bnew+ker}
\ker\,(\mathcal{B}^+)=\text{span}\,\{
\Psi_0,A^-\psi^-_0,A^-\psi^-_1,A^-\psi^-_3,A^-\widetilde{\psi^-_1}\}\,.
\ee
The eigenvalue $E=2$ of the 
non-physical eigenstates 
$A^-\psi^-_1$ and $A^-\widetilde{\psi^-_1}$ 
in the kernel of the increasing  ladder operator 
$\mathcal{B}^+$
corresponds to the
double root of the last factor in the 
second relation in  
(\ref{B+newB-}). 
By analogy with (\ref{laddern+1}), 
one can consider the 
higher-order ladder operators
\be
\mathcal{B}^\pm_{n}=\B^-_2(a^\pm)^n\B^-_2\,,
\ee
$[\breve{H},\mathcal{B}^\pm_{n}]=\pm 2n
\mathcal{B}^\pm_{n}$,
with the identification $\mathcal{B}^\pm_1=\mathcal{B}^\pm$.
The  lowering operator $\mathcal{B}^-$  can be   related 
to the ladder operators $\mathcal{A}^-$
and $\mathcal{C}^-$
via the identities 
\be\label{BnACH-4}
\mathcal{B}^-=\mathcal{A}^-(\breve{H}-4)\,,\qquad
\mathcal{B}^-_2=(\mathcal{A}^-)^2\,,\qquad
\mathcal{B}^-_3 =-\mathcal{C}^-(\breve{H}-2)(\breve{H}-4)\,,
\ee
$\mathcal{B}^-_4=-\mathcal{C}^-_2(\breve{H}-2)(\breve{H}-4)$,
$\mathcal{B}^-_5=-\mathcal{C}^-(\mathcal{A}^-)^2$,
$\mathcal{B}^-_6=(\mathcal{C}^-)^2(\breve{H}-2)(\breve{H}-4)$,
etc.  The  increasing operator $\mathcal{C}^+$  is related 
to  $\mathcal{A}^+$
and $\mathcal{C}^+$ via   the conjugate identities.
Similarly to (\ref{A-n=An-}), for degrees $n>1$ 
of $\mathcal{B}^-$ we have 
\be
(\mathcal{B}^-)^n=\prod_{j=1}^{n-1}(\breve{H}-2+2j)
(\breve{H}-4+2j)\cdot \mathcal{B}^-_n\,,
\ee
and an analogous  relation for $(\mathcal{B}^+)^n$.

 A generalization of the $(\beta_2)$-scheme 
  corresponds to the family of the DCKAT schemes
\be
\quad (\beta_3)=\{\psi_0,\psi_2,\psi_3\}\,, \ldots,
(\beta_{n+2})=\{\psi_0,\ldots,\psi_{n-1},\psi_{n+1},\psi_{n+2}\}\,.
\ee
In the case of the scheme 
$(\beta_{n+2})$, the intertwining operators  
constructed according to the prescription (\ref{defAn}) 
are $\B_{n+2}^-\equiv \B^-_2(a^-)^n$
and $\B^+_{n+2}=(a^+)^n\B_2^+$.
The   schemes $(\beta_{n+2})$ 
do not give anything new but 
allow us to re-interprete 
the already discussed higher-order
ladder operators $\mathcal{C}^\pm_{n+1}$
as those produced
via the composition of the $(\beta_{n+2})$ and $(\alpha_1)$ schemes,
$\mathcal{C}^-_{n+1}=\B^-_{n+2}A^+$, 
$\mathcal{C}^+_{n+1}=A^-\B^+_{n+2}$.
Analogously, $\mathcal{B}^-_{n+1}=\B^-_{n+2}\B^+_2$,
$\mathcal{B}^+_{n+1}=\B^-_2\B^+_{n+2}$.

Besides the two infinite families $(\alpha_n)$,
$n=1,\ldots$,  and $(\beta_n)$, $n=2,\ldots$,
which involve  as the seed eigenfunctions
 either only  non-physical 
or only physical eigenstates of the QHO,
there are two additional, `intermediate' schemes 
which simultaneously include  eigenstates  of   
both types. These are 
the schemes
\be
(\gamma_2)=\{ \psi_0,\psi^-_1\}\,,\qquad
(\gamma_3)=\{\psi_0, \psi_1,\psi^-_0\}\,.
\ee
The intertwining operators 
in the case of the scheme $(\gamma_2)$
are the second-order differential operators
$\A_2=A^-a^-$ and $\A_2^\dagger=a^+A^+$, while 
in the scheme  $(\gamma_3)$,
the intertwining operators are 
the third-order  differential operators
$\A_3=A^-(a^-)^2$ and $\A_3^\dagger=(a^+)^2A^+$.
These operators have a structure similar 
to that of  the intertwining operators in 
the family of the schemes $(\alpha_n)$.
With their help we do not 
obtain anything essentially new for the construction 
of the ladder operators for the REQHO system
$\breve{H}$ in comparison 
with the already discussed structures.
Indeed, employing the superposition 
of the intertwining operators from the $(\beta_2)$-scheme
and either $(\gamma_2)$- or $(\gamma_3)$-  schemes, 
one can construct the ladder operators 
\be\label{B-1B-2}
\mathcal{C}^-_{-n}\equiv \B^-_2(a^+)^n A^+\,,\qquad 
\mathcal{C}^+_{-n}\equiv A^- (a^-)^n  \B^+_2\,,\qquad
n=1,2\,.
\ee
Here  $n=1$ and $n=2$ correspond, respectively,
to the $(\gamma_2)$- and  $(\gamma_3)$- schemes.
Operators $\mathcal{C}^\pm_{-1}$ are 
fourth-order differential operators,  
while $\mathcal{C}^\pm_{-2}$ are  fifth-order differential 
operators.
They, however,  are not independent 
but  can be expressed in terms of 
the already constructed intertwining operators.
Namely, we have, in particular,  
\be\label{B-ndef}
\mathcal{C}^\pm_{-1}=-\mathcal{A}^\pm_2\,,\qquad
\mathcal{C}^-_{-2}=-(\breve{H}-2)\mathcal{A}^-\,,\qquad
\mathcal{C}^+_{-2}=-(\breve{H}-4)\mathcal{A}^+\,.
\ee
These relations can be established by  comparing  the  
kernels of  $\mathcal{C}^-_{-1}$ and 
$\mathcal{C}^-_{-2}$,
$\ker\, (\mathcal{C}^-_{-1})=\text{span}\,
\{\Psi_0,A^-\psi^-_0,\Psi_1,\Psi_2\}$, 
$\ker\, (\mathcal{C}^-_{-2})=\text{span}\,
\{\Psi_0,A^-\psi^-_1, A^-\psi^-_0,
A^-\widetilde{\psi^-_0}, 
\Psi_1\}$,
with the kernels of 
$\mathcal{A}^-_2$
and of the operator 
$(\breve{H}-2)\mathcal{A}^-=\mathcal{A}^-(\breve{H}-4)$, respectively,
and by comparison of the signs before the leading 
derivative terms in the corresponding pairs of
operators.
Due to the identities  (\ref{ABa3}), 
a generalization of the operators (\ref{B-1B-2})
for $n>2$  does not give us 
anything new  since 
$\mathcal{C}^\pm_{-3}=-\breve{H}(\breve{H}-2)
(\breve{H}-4)$.
This last relation as well as relations 
(\ref{B-ndef}) can also be obtained by
employing the identities (\ref{ABa3}).
We also  have the operator identities 
\be
\mathcal{A}^+\mathcal{C}^+_n=(\breve{H}-2)\mathcal{C}^+_{n+1}\,,\qquad
\mathcal{A}^-\mathcal{C}^-_n=-\mathcal{A}^-_{n+3}\,,\qquad
n=1,\ldots,
\ee
\be
\mathcal{A}^+\mathcal{C}^-=-(\breve{H}-6)\mathcal{A}^-_{2}\,,\qquad
\mathcal{A}^-\mathcal{C}^+=-(\breve{H}+2)\mathcal{A}^+_{2}\,,
\ee
as well as the Hermitian conjugate 
relations.

Let us look  in more detail at 
the already 
mentioned similarity between the operators
$\mathcal{C}^\pm$ and $\mathcal{A}_3^\pm$.
Using the first identity from (\ref{ABa3}) and 
factorization relation (\ref{breveHdef}), we obtain 
 $\mathcal{A}^-_{3}=A^-(a^-)^3A^+
=-\breve{H}\mathcal{C}^-$ and 
$\mathcal{A}^+_{3}=-(\breve{H}-6)\mathcal{C}^+$.
These relations are similar to those in (\ref{B-ndef}).
Employing Eq. (\ref{A-n=An-}),
we find  that 
\be\label{A-3HB-}
(\mathcal{A}^-)^3=-\breve{H}
(\breve{H}+2)(\breve{H}+4)\mathcal{C}^-=
-\mathcal{C}^-(\breve{H}-6)
(\breve{H}-4)(\breve{H}-2)\,.
\ee
This relation from the point of view 
of the kernels of the involved 
operators  corresponds to the following picture.
The kernel of the operator $\mathcal{A}^-$ 
is formed by  the two physical eigenstates
$\Psi_0$ and $\Psi_1$
and by one non-physical eigenstate $A^-\psi^-_1$.
We have also the relations \cite{CarPly}
$\mathcal{A}^-(A^-\widetilde{\psi^-_1})=\Psi_0$,
$\mathcal{A}^-(A^-\psi^-_0)=A^-\psi^-_1$,
$\mathcal{A}^-\Psi_2=\Psi_1$,
$\mathcal{A}^-(A^-\widetilde{\psi^-_0})=A^-\widetilde{\psi^-_1}$,
$\mathcal{A}^-(A^-\widetilde{\psi_0})=A^-\psi^-_0$,
$\mathcal{A}^-\Psi_3=\Psi_2$. As a result we obtain
\be\label{kerA3}
\ker\,(\mathcal{A}^-)^3=\text{span}\,\{
\Psi_0,
A^-\psi^-_1,
 \Psi_1, 
 A^-\widetilde{\psi^-_1},
A^-\psi^-_0,
 \Psi_2, 
 A^-\widetilde{\psi^-_0},
A^-\widetilde{\psi_0},\Psi_3\}\,.
\ee
The pairs  of states ($A^-\psi^-_1$, 
$A^-\widetilde{\psi^-_1}$), 
($A^-\psi^-_0$, $\widetilde{\psi^-_0}$)
and ($\Psi_1=A^-\psi_0$, $A^-\widetilde{\psi_0}$)
constitute, respectively,   
the kernels of the factors $(H-2)$,
$(H-4)$ and $(H-6)$
in (\ref{A-3HB-}).
The remaining three physical eigenstates  
$\Psi_0$, $\Psi_2$ and $\Psi_3$ in (\ref{kerA3})
correspond to the kernel of the 
operator $\mathcal{C}^-$.

According to  Eq. (\ref{A-3HB-}) 
and its conjugate version,  the ladder operators
$\mathcal{C}^\pm$ can be generated 
by $\mathcal{A}^\pm$.
Then with taking into account Eq. (\ref{BnACH-4})
and all the described relations,  
we conclude that in the case of the simplest 
REQHO system given by the Hamiltonian
$\breve{H}$ defined in (\ref{breveHdef})
all the set of the  ladder operators
can be obtained, in principle, 
from the  compositions of  
the 
ladder
operators $\mathcal{B}^+$
and  $\mathcal{B}^-$.

In conclusion of this section 
let us note, however,  that in comparison with the 
QHO picture, the peculiarity of the system
(\ref{breveHdef}) in particular is that
its  ground-state $\Psi_0$ 
cannot be achieved 
from physical states
by action of
the lowering  operators $\mathcal{A}^-$ and $\mathcal{B}^-$
which are  differential operators of orders $3$ and $5$.  
Like the first-order differential operator
$a^-$ in the QHO,  the ladder operators
 $\mathcal{A}^-$ and $\mathcal{B}^-$ 
decrease the energy values of $\breve{H}$ 
 in $2$, but they produce 
the ground-state by acting on the 
non-physical eigenstate $A^-\widetilde{\psi^-_1}$
of the eigenvalue $E=2$.
The ground-state $\Psi_0$ of zero energy can be achieved,
however,
by application of the lowering operator
$\mathcal{C}^-$, which is a third-order differential operator,
  to the physical eigenstate
$\Psi_1$  with eigenvalue $E=6$.

We also notice here  that the ladder operators 
(\ref{ladderA1}) for the REQHO system 
of the simplest form  (\ref{breveHdef}) were constructed 
(without  employing   the Darboux-dressing procedure)  in  
 \cite{Dub} where this model was introduced and 
 investigated  for the first time.
Later these ladder operators were constructed,
particularly,  in \cite{JunRoy},
\cite{MarQue1+,MarQue1} and recently in \cite{CarPly}
via the  Darboux-dressing prescription 
based on the non-physical seed state
we used here.
The ladder operators (\ref{CB2A}) we constructed
on the basis of the $(\alpha_1)$- and $(\beta_2)$- schemes 
by employing the analogy with  reflectionless 
quantum systems \cite{CoJaPl,LaxNov}
where the corresponding Lax-Novikov integrals of motion can be
generated either by Darboux-dressing of the
free particle momentum operator or by `gluing'
two different intertwining 
operators that act in the opposite directions. 
The same last mentioned method 
also allows ones to generate the Lax-Novikov integrals 
for periodic finite-gap systems where
the Darboux-dressing mechanism  can not be applied, see  
\cite{exosusy}. 
In reflectionless and finite-gap systems, 
however, the 
two glued  intertwiners always 
are differential operators of the 
`opposite', even and odd, differential orders,
but both intertwine the two corresponding 
partner systems
without additional relative displacement.
It is because of the relative  displacement 
in intertwining relations (\ref{AHinter}) 
and  (\ref{B2+6}) that here we obtain the
ladder operators $\mathcal{C}^\pm$ 
while in reflectionless  and finite-gap systems
analogous procedure generates the integrals 
of motion.
Within the same framework we used here and 
based on  employing the
two schemes with physical and non-physical seed states
of the quantum harmonic oscillator,  
the ladder operators  (\ref{CB2A}) were introduced
earlier in \cite{MarQue1} 
 (but without exploiting the 
indicated analogy with generation of the Lax-Novikov integrals)
and derived later in \cite{MarQue1++} 
for  more general families related to 
multi indexed exceptionnal orthogonal polynomials.
By another method such ladder operators were
introduced  for the system (\ref{breveHdef}) 
even earlier in \cite{Spir}. 
The ladder operators (\ref{B2aB2})  constructed here
by the Darboux-dressing procedure based on physical seed states 
seems were not discussed earlier in the literature.

With subsequent  analysis we shall see that the 
trinity 
of the basic ladder operators ($\mathcal{A}^\pm$,
$\mathcal{B}^\pm$, $\mathcal{C}^\pm$) 
admits a natural generalization for the case of
 REQHO systems of a general form, and 
that each pair of the conjugate lowering and raising 
ladder operators  
detects and reflects some specific properties 
of a corresponding quantum system.

\section{Two further examples of the REQHO systems}\label{sec+2systems}

A  REQHO system generated by 
the  DT
based on the non-physical 
state $\psi^-_{2n}$ with $n>1$ is 
similar to the considered 
REQHO system generated by
the DT based on $\psi^-_2$.
In this case the gap in the spectrum 
of the REQHO corresponds,
up to a global shift,  to the missing 
$2n$ energy levels  with  $E=3,\ldots,4n+1$
in the spectrum of the QHO. 
We also have here the two infinite families of
the DCKAT schemes
of the structures which generalize those of the 
case $n=1$.  For instance, in the case of $n=2$, 
we have $\psi^-_4=(4x^4+12x^2+3)e^{x^2/2}$, and
the two infinite families of the schemes are
\be\label{alpha4n}
(\alpha_1)=\{\psi^-_4\}\,,\qquad
(\alpha_2)=\{\psi^-_0,\psi^-_5\}\,,\qquad
(\alpha_{n+1})=\{\psi^-_0,\ldots,\psi^-_{n-1},\psi^-_{n+4}\}\,,
\ee
and
\be\label{beta4n}
(\beta_4)=\{\psi_1,\psi_2,\psi_3,\psi_4\}\,,\qquad
(\beta_{n+4})=\{\psi_0,\ldots,\psi_{n-1},
 \psi_{n+1},\psi_{n+2},\psi_{n+3},\psi_{n+4}\}\,.
 \ee
 In addition, we have the `intermediate' schemes whose sets 
 of seed states include both physical and non-physical 
 eigenstates of the QHO.
 These are
 \be\label{gamma4n}
 (\gamma_2)=\{\psi_0,\psi^-_3\}\,,
 \quad
  (\gamma_3)=\{\psi_0,\psi_1,\psi^-_2\}\,,
 \quad
 (\gamma_4)=\{\psi_0,\psi_1,\psi_2,\psi^-_1\}\,,
 \quad
 (\gamma_5)=\{\psi_0,\psi_1,\psi_2,\psi_3,\psi^-_0\}\,.
  \ee
The scheme $(\alpha_1)$  generates
the intertwining operators $A^-=\psi^-_4\frac{d}{dx}\frac{1}{\psi^-_4}$,
and $A^+=(A^-)^\dagger$.
They allow us to construct ladder operators that are third-order differential
operators,  the Darboux-dressed 
ladder operators of the QHO, $\mathcal{A}^\pm= A^-a^\pm A^+$.
They satisfy the relations of the form 
(\ref{HcomA}), 
$[\breve{H},\mathcal{A}^\pm]=\pm 2\mathcal{A}^\pm$,   with
\be\label{HAA32}
\breve{H}\equiv A^-A^+=
-\frac{d^2}{dx^2}+x^2+7+32\frac{4x^6+4x^4+3x^2-6}{(4x^4+12x^2+3)^2}\,,
\ee
and similarly to the 
already considered case,
here both ladder operators $\mathcal{A}^\pm$ annihilate the
ground state of $\breve{H}$ which is $\Psi_0=1/\psi_4^-=A^-\widetilde{\psi^-_4}$.
Besides the ground state $\Psi_0$ 
of  energy $E=0$,  the kernel of $\mathcal{A}^-$ contains, 
the first excited state $\Psi_1=A^-\psi_0$ of  
energy $E=10$
and one non-physical eigenstate $A^-\psi^-_3$ of 
energy $E=2$.
The kernel of $\mathcal{A}^+$ contains besides the ground state 
$\Psi_0$ also two non-physical eigenstates 
$A^-\psi^-_0$ and $A^-\psi^-_5$
of $\breve{H}$ of eigenvalues $E= 8$ and $E=-2$. 
The intertwining operators corresponding to the 
DCKAT scheme $(\beta_4)=\{\psi_1,\psi_2,\psi_3,\psi_4\}$ are the 
fourth order differential operators
constructed in accordance with 
Eq. (\ref{defAn}),
which by analogy with the already 
considered case 
we denote here as
$\B^-_4$ and $\B^+_4=(\B^-_4)^\dagger$.
Then we define another pair of   ladder operators
via a composition of the intertwining  operators
of this $(\beta_4)$-scheme and of the 
$(\alpha_1)$-scheme,
$\mathcal{C}^-=\B^-_4 A^+$, $\mathcal{C}^+=
A^-\B^+_4$.
Unlike the previously discussed case of the REQHO system
(\ref{breveHdef}),
these are fifth-order differential operators,
which satisfy the relations $[\breve{H},\mathcal{C}\pm]=
\pm 10\mathcal{C}^\pm$.
The ladder operator $\mathcal{C}^-$ annihilates
five physical eigenstates of $\breve{H}$,
which are the ground state 
$\Psi_0$ and the states $\Psi_{j+1}=A^-\psi_j$, $j=1,2,3,4$,
with the  energy values $E_0=0$ and
$E_{j+1}=10+2j$.
Like in the REQHO system we considered before, 
 the first excited state $\Psi_1=A^-\psi_0$ of energy
$E_1=10$ here does not belong to the kernel of 
the decreasing ladder operator $\mathcal{C}^-$,
and we have $\mathcal{C}^-\Psi_1=\psi_0$, 
$(\mathcal{C}^-)^2\Psi_1=0$.
The kernel of $\mathcal{C}^+$
is composed only from non-physical eigenstates.
Yet another pair of the ladder operators corresponds to
$
\mathcal{B}^\pm=\B^-_4a^\pm\B^+_4
$,
 which are differential operators of order $9$.
 Like $\mathcal{A}^\pm$, they satisfy the 
 relations 
 $[\breve{H},\mathcal{B}^\pm]=\pm 2\mathcal{B}^\pm$.
 The kernel of the  lowering ladder operator 
 $\mathcal{B}^-$ is spanned by two physical eigenstates 
 $\Psi_0$ and $\Psi_1$
 of energies $0$ and $10$, and seven non-physical eigenstates
 of $\breve{H}$ of eigenvalues $8$ (twice), $6$ (twice), $4$ (twice)
 and  $2$,
  $\ker\,(\mathcal{B}^-)=\text{span}\, \{
 \Psi_0,\Psi_1,A^-\psi^-_0,
 A^-\widetilde{\psi^-_0},\Psi_1,A^-\psi^-_1,
 A^-\widetilde{\psi^-_1},
 \Psi_1,A^-\psi^-_2,
 A^-\widetilde{\psi^-_2},A^-\psi^-_3\}$.  
 The kernel of $\mathcal{B}^+$ is spanned by the 
 ground-state $\Psi_0$ and by eight non-physical eigenstates
 of $\breve{H}$.

Other,  
secondary ladder  operators can be constructed 
by introducing the QHO ladder operators $(a^\pm)^n$
inside the factorized structures 
of the basic ladder operators
$\mathcal{A}^\pm$, $\mathcal{B}^\pm$
and $\mathcal{C}^\pm$,
or by considering compositions
of the intertwining operators corresponding to
(\ref{alpha4n}), (\ref{beta4n}) and (\ref{gamma4n}) schemes 
analogously to
 how it was done for the simplest REQHO system.
 The  secondary, higher-order ladder operators can also be
 generated by taking the products of  the 
 basic ladder operators.
 Analogously to (\ref{ABa3}), we also have here 
 the relations 
 $A^+\B^-_4=-(a^-)^5$, 
$\B^+_4A^-=-(a^+)^5$.
As an analog of relation (\ref{A-3HB-})
we have 
\be\label{A5HB-}
(\mathcal{A}^-)^5=
-\mathcal{C}^-(\breve{H}-10)
(\breve{H}-8)(\breve{H}-6)(\breve{H}-4)\,,
\ee
and the relation 
\be\label{B-A-H2H4H6}
\mathcal{B}^-=(\breve{H}-2)(\breve{H}-4)(\breve{H}-6)\mathcal{A}^-
\ee
is analogous here to the first relation in (\ref{BnACH-4}). 
As in the case of the simplest REQHO system
considered in the previous section,
for the  REQHO system described by the Hamiltonian(\ref{HAA32})  
all the ladder operators can be generated and extracted 
from the powers of the basic ladder operators $\mathcal{B}^\pm$.
 \vskip0.1cm
 
Let us consider yet  another  example of the REQHO system
in which  two states are separated by a gap 
from the infinite equidistant part of the spectrum.
A simplest system of such a nature
 can be
 generated by employing  the  minimal $(\alpha)$-scheme 
\be\label{alpha2_2_3}
(\alpha_2)=\{\psi^-_2,\psi^-_3\}\,.
\ee
Let us shift the Hamiltonian of the QHO for $+7$ 
and denote  $H=H_{\textrm{osc}}+7$,  for which the potential is 
$V(x)=x^2+7$ and the spectrum is $E_n=8+2n$, $n=0,1,\ldots$.
The Wronskian here is
 $\W_2(x)=e^{x^2}(3+4x^4)$.
The second order DCKAT based on (\ref{alpha2_2_3}) produces 
the partner system which is  the REQHO system 
$\breve{H}=-\frac{d^2}{dx^2}+\breve{V}(x)$,
where in accordance with (\ref{VnWr}),
\be\label{potentialx4}
\breve{V}(x)=3+x^2
+32x^2\frac{4x^4-9}{(3+4x^4)^2}\,.
\ee
Its gapped spectrum is 
\be 
E_0=0\,,\quad 
E_1=2\,,\quad 
E_{2+n}=8+2n\,,\quad 
n=0,1,\ldots\,.
\ee
The intertwining second order differential operators
constructed via (\ref{defAn}) on the basis of the seed 
QHO eigenstates (\ref{alpha2_2_3}) 
 we denote as $\A_2^-$ and
$\A_2^+=(\A_2^-)^\dagger$.
The kernel of $\A^-_2$ is spanned by the states
$\psi^-_2$ and $\psi^-_3$, 
while the kernel of $\A^+_2$ is spanned by 
the lowest physical eigenstates of $\breve{H}$
of the energies $0$ and $2$,
which can be obtained from 
the QHO non-physical eigenstates $\widetilde{\psi^-_2}$
and $\widetilde{\psi^-_3}$,
$\Psi_0=\A^-_2\widetilde{\psi^-_2}$,
$\Psi_1=\A^-_2\widetilde{\psi^-_3}$.
The operators $\A^+_2$ and $\A^-_2$ 
satisfy the relations
\be
\A^+_2\A^-_2=H(H-2)\,,\qquad 
\A_2^-\A^+_2=\breve{H}(\breve{H}-2)\,,
\ee
and
$
\A^-_2H=\breve{H}\A^-_2$,
$\A^+_2\breve{H}=H\A_2^+$.
We construct the ladder operators $\mathcal{A}^\pm$  for $\breve{H}$
by the Darboux-dressing of the QHO operators $a^\pm$,
\be
\mathcal{A}^\pm =
\A^-_2 a^\pm \A_2^+\,.
\ee
These fifth-order differential operators satisfy 
the relations 
$
[\breve{H},\mathcal{A}^\pm]=\pm 2\mathcal{A}^\pm
$,
and 
\be\label{polynom+2}
\mathcal{A}^+\mathcal{A}^-=\breve{H}(\breve{H}-2)^2(\breve{H}-4)(\breve{H}-8)\,,\qquad
\mathcal{A}^-\mathcal{A}^+=(\breve{H}+2)(\breve{H})^2(\breve{H}-2)(\breve{H}-6)\,.
\ee
The kernel of the lowering operaror is
$\ker\,(\mathcal{A}^-)=\text{span}\,\{\Psi_0,\Psi_1,\Psi_2, \widetilde{\Psi_1}, \A_2^-\psi^-_1\}$.
Here the first three states are the lowest three 
physical eigenstates of $\breve{H}$ of 
energies $E=0$, $2$ and $8$,
respectively, and the two last states are non-physical eigenstates 
of energies $E=2$ and $4$. 
The indicated energies correspond 
to the roots of the polynomial in the first equality in (\ref{polynom+2}).
The kernel of the increasing operator 
$\mathcal{A}^+$ is spanned by the two lowest eigenstates
$\Psi_0$ and $\Psi_1$ and by the three non-physical eigenstates:
$\ker\,(\mathcal{A}^+)=
\text{span}\,\{\Psi_0,\Psi_1,\A^-_2\psi^-_4,\widetilde{\Psi_0},\A^-_2\psi^-_0\}$.
The energies of these five eigenstates 
correspond to zeros of the polynomial in 
the second equality in  (\ref{polynom+2}).

The same system, up to a global shift, 
can also be generated 
via  the complementary minimal $(\beta_2)$-scheme
\be\label{seedbeta2}
(\beta_2)=\{\psi_2,\psi_3\}\,.
\ee
The Wronskian of the seed states in this case 
is 
$\W_2(x)=\W(\psi_2,\psi_3)=
e^{-x^2}(3+4x^4)$.
The potential calculated according to (\ref{VnWr})
is the potential (\ref{potentialx4}) shifted for $+8$.
Let us denote the corresponding intertwining second order
differential operators constructed on the basis of these seed states
as $\B^-_2$ and $\B^+_2=(\B^-_2)^\dagger$.
They satisfy the relations 
 \be
 \B^-_2H=(\breve{H}+8)\B^-_2\,,\qquad
 \B^+_2(\breve{H}+8)=H\B^+_2\,,
 \ee
 and
 \be
\B^+_2\B^-_2=(H-12)(H-14)\,,\qquad
\B^-_2\B^+_2=(\breve{H}-4)(\breve{H}-6)\,.
\ee
Note that here 
$\B_2^-(x) =-\A_2^-(ix)$.
The kernel of $\B^-_2$ is spanned by the seed states
(\ref{seedbeta2}), whereas the kernel of $\B^+_2$ 
is spanned by non-physical 
eigenstates of $\breve{H}$ of energies
$E=4$ and $E=6$:
$\ker\, (\B^+_2)=\text{span}\,\{\B^-_2\widetilde{\psi_2},\B^-_2\widetilde{\psi_3}\}=
\text{span}\,\{\A^-_2\psi^-_0,\A^-_2\psi^-_1\}$.
The ladder operators 
\be\label{B4ladder}
\mathcal{C}^-=\B^-_2\A^+_2\,,\qquad
\mathcal{C}^+=\A^-_2\B^+_2
\ee
 are differential 
operators of the order four.
They  obey the relations 
\be
[\breve{H},\mathcal{C}^\pm]=\pm 8\,\mathcal{C}^\pm\,,
\ee
and 
\be
\mathcal{C}^+\mathcal{C}^-=\breve{H}(\breve{H}-2)
(\breve{H}-12)(\breve{H}-14)\,,\qquad
\mathcal{C}^-\mathcal{C}^+=(\breve{H}+8)(\breve{H}+6)
(\breve{H}-4)(\breve{H}-6)\,.
\ee
The lowering operator $\mathcal{C}^-$ annihilates
the four physical states $\Psi_0,\Psi_1,\Psi_4,\Psi_5$
of the energies
$E=0,2,12,14$.
The kernel of $\mathcal{C}^+$ is spanned only by non-physical 
eigenstates 
of energies $E=-8,-6,4,6$,
$\ker\,(\mathcal{C}^+)=\text{span}\,
\{\B^-_2\psi^-_3,\B^-_2\psi^-_2,\B^-_2\widetilde{\psi_2},
\B^-_2\widetilde{\psi_3}\}=
\text{span}\,
\{\A^-_2\psi^-_7,\A^-_2\psi^-_6,\A^-_2\psi^-_1,
\A^-_2\psi^-_0\}$.
Changing the order of the second order operators 
in the factorized form of the ladder operators in 
(\ref{B4ladder}),  we 
obtain the operator identities
\be\label{B2A2a}
\B^+_2\A^-_2=(a^+)^4\,,\qquad
\A^+_2\B^-_2=(a^-)^4\,,
\ee
cf. (\ref{ABa3}).
Again, these relations  can be verified   by comparing the 
kernels of the corresponding operators.

Yet another pair of the ladder operators corresponds to 
differential operators of order $5$,
\be
\mathcal{B}^\pm =
\B^-_2 a^\pm \B_2^+\,.
\ee
They satisfy relations $[\breve{H},\mathcal{B}^\pm]=
\pm 2\mathcal{B}^\pm $, and 
\be\label{polynom+2B}
\mathcal{B}^+\mathcal{B}^-=\breve{H}(\breve{H}-4)(\breve{H}-6)^2(\breve{H}-8)\,,\qquad
\mathcal{B}^-\mathcal{B}^+=(\breve{H}+2)(\breve{H}-2)(\breve{H}-4)^2(\breve{H}-6)\,.
\ee
In correspondence with the first relation in (\ref{polynom+2B}), 
the kernel of the lowering operator is spanned 
by the ground-state $\Psi_0$ and the eigenstate $\Psi_2$ at the bottom 
of the equidistant infinite part of the spectrum as well as by 
the three non-physical eigenstates
of $\breve{H}$,
$\ker\,(\mathcal{B}^-)=\text{span}\,\{\Psi_0,\Psi_2, \A_2^-\psi^-_0,
\A_2^-\psi^-_1,\A_2^-\widetilde{\psi^-_0}\}$.
The kernel of the increasing operator $\mathcal{B}^+$
is spanned by the first excited physical eigenstate
$\Psi_1$ of energy $2$, and by four non-physical eigenstates 
of $\breve{H}$ of eigenvalues  $-2$, $4$ (twice) and $6$,
$\ker\,(\mathcal{B}^+)=\text{span}\,\{\Psi_1,\A_2^-\psi^-_3,
\A_2^-\psi^-_1,\A_2^-\widetilde{\psi^-_1},\A_2^-\psi^-_0\}$.
Here $\mathcal{B}^-$ and  $\mathcal{A}^-$
are related by the operator identity
\be\label{BAn+=2n-=1}
\breve{H}\mathcal{B}^-=(\breve{H}-4)\mathcal{A}^-\,,
\ee
and the relation between the increasing ladder operators 
 $\mathcal{B}^+$ and  $\mathcal{A}^+$
 is given by Hermitian conjugation of 
 (\ref{BAn+=2n-=1}). 
 Note that in comparison with the first relation 
 in (\ref{BnACH-4}) and relation (\ref{B-A-H2H4H6})
 in the examples of the REQHO systems 
 with one separated energy level here 
 the relation (\ref{BAn+=2n-=1}) contains 
 a Hamiltonian-dependent factor before the operator 
 $\mathcal{B}^-$.  Coherently with this,
 in the REQHO system under consideration 
 the lowering operator $\mathcal{B}^-$ in comparison 
 with $\mathcal{A}^-$ annihilates only 
 the lowest state $\Psi_0$ 
  in the separated part of the spectrum.
 The raising operator $\mathcal{B}^+$ annihilates 
 another separated state $\Psi_1$ in comparison 
 with both separated states $\Psi_0$ and $\Psi_1$ 
 annihilated by $\mathcal{A}^+$.
 This difference can be understood  if we 
 note that the polynomial in the first identity in (\ref{polynom+2B})
 does not have the root $2$.
 Rewriting  relation (\ref{BAn+=2n-=1}) in the equivalent form
 $\mathcal{B}^-(\breve{H}-2)=\mathcal{A}^-(\breve{H}-6)$,
 we see then that the annihilation of the state
 $\Psi_1$  from the kernel of $\mathcal{A}^-$
 is provided by the factor $(\breve{H}-2)$ on the left hand side 
 of the identity. In the same way one can understand 
 the difference in the kernels of the raising operators
  $\mathcal{B}^+$ and $\mathcal{A}^+$ by looking at 
  the roots of the polynomial in the second identity 
  in (\ref{polynom+2B}) and by taking into account 
  the identity relation 
  $\mathcal{B}^+\breve{H}=\mathcal{A}^+(\breve{H}-4)$
  to be conjugate to (\ref{BAn+=2n-=1}).

The same (up to a global shift) system $\breve{H}$ given by 
the potential (\ref{potentialx4}) can be produced by using the
higher-order  $(\alpha)$- and $(\beta)$-schemes,
\be
(\alpha_{n+2})=\{\psi^-_0,\ldots,\psi^-_{n-1},\psi^-_{n+2},\psi^-_{n+3}\}\,,\qquad
(\beta_{n+2})=\{\psi_0,\ldots,\psi_{n-1},\psi_{n+2},\psi_{n+3}\}\,,
\ee
$n=1,\ldots$, 
and by the  two intermediate 
$(\gamma)$-schemes,
\be
(\gamma_3)=\{\psi_0,\psi^-_1,\psi^-_2\}\,,\qquad
(\gamma_4)=\{\psi_0,\psi_1,\psi^-_0,\psi^-_1\}\,.
\ee
The secondary, higher-order ladder operators can be generated here
in the same way as  for the REQHO systems
with one added gapped bound state.
Here the relation
\be\label{A4HB-}
(\mathcal{A}^-)^4=
\mathcal{C}^-(\breve{H}-8)^2
(\breve{H}-6)^2(\breve{H}-4)^2(\breve{H}-2)(\breve{H}-10)\,.
\ee
is analogous to the  relations (\ref{A-3HB-}) and 
(\ref{A5HB-}),
 and shows that the ladder operators
$\mathcal{C}^\pm$ can be  generated by
the  ladder operators $\mathcal{A}^\pm$.

\section{REQHO systems of a general form and their ladder operators}\label{sec:general}

We generalize now our analysis  of the particular examples
for the case of  REQHO systems 
of a  general form, for which 
we construct    the ladder operators and investigate their
properties.

Each REQHO system can be generated 
by employing 
the appropriate DCKAT based on 
any of the schemes from the two infinite families.
The $(\alpha)$-type schemes
include only 
non-physical eigenstates of the QHO   
chosen as
the seed states.
The $(\beta)$-type schemes involve 
only the corresponding physical eigenstates
of the QHO.
Besides, there exists also a  finite number
of intermediate $(\gamma)$-type schemes
which simultaneously 
use the eigenstates of both types.
Let us denote $\A_{n_+}^-$ and $\A_{n_+}^+$
 the mutually conjugate intertwining operators 
constructed on the basis of the $(\alpha)$-type scheme with
a minimal number  $n_+$ of  seed non-physical eigenstates. 
They are
 differential operators of order $n_+$.
Analogously, let us denote the 
intertwining operators constructed on the basis
of the $(\beta)$-scheme with a minimal number $2n_-$
of seed physical eigenstates as 
$\B_{2n_-}^-$ and $\B_{2n_-}^+$.
By the construction we have 
$\A^-_{n_+}\psi^-_{j_s}=0$ and $\B^-_{2n_-}\psi_{i_s}=0$,
where $\psi^-_{j_s}$ are $n_+$ non-physical eigenstates 
of the QHO
which are the seed states in the  $(\alpha_{n_+})$-scheme,
 $\psi^-_{j_s}\in (\alpha_{n_+})$, while
 $\psi_{i_s}$ are $2n_-$  physical eigenstates which are used 
 in the $(\beta_{2n_-})$-scheme, 
 $\psi_{i_s}\in (\beta_{2n_-})$. 
 Such two minimal $(\alpha_{n_+})$-
and $(\beta_{2n_-})$-schemes are complementary 
similarly to the  schemes $(\alpha_1)$ and $(\beta_2)$
in the case of the simplest REQHO system
we considered in detail above.
 The corresponding 
Wronskians in these two schemes  have the form 
$\W_{n_+}=\exp{(\frac{1}{2}x^2n_+)}\phi(x)$
and 
$\W_{2n_-}=c\exp{(-x^2 n_-)}\phi(x)$,
where $\phi(x)$ is some nodeless 
polynomial 
function, 
and  $c$ is some constant.
We fix the additive constant shifts 
in  the Hamiltonians $H$ of the QHO  
and $\breve{H}$ of the REQHO 
in such a way that $\A^-_{n_+}H=\breve{H}\A^-_{n_+}$
and  that the ground-state $\Psi_0$ of $\breve{H}$ 
has 
zero  energy, $E_0=0$.  Then 
the lowest state $\Psi_{n_+}=\A^-_{n_+}\psi_0$ in the equidistant  
infinite part of the spectrum of $\breve{H}$ will be characterized
by the energy value 
\be\label{En+n-def}
E_{n_+}=2(n_++2n_-)\equiv 2\Delta
\ee
that also will be the energy of the ground-state $\psi_0$ of 
the shifted QHO Hamiltonian $H$,
$H\psi_0=E_{n_+}\psi_0$.
Then we have $H=H_{\textrm{osc}}-1+2\Delta$.
For the other basic lowering 
intertwining operator $\B^-_{2n_-}$
we have the relation
 \be\label{BHbrHE}
 \B^-_{2n_-}H=(\breve{H}+2\Delta)\B^-_{2n_-}\,.
 \ee
This  means in particular that if
 $\psi(x;E)$ is an eigenstate of $H$ of energy $E$ and 
 if $\B^-_{2n_-}\psi(x;E)\neq 0$,
 then this latter state  will be
 eigenstate of $\breve{H}$
 of the eigenvalue $(E-2\Delta)$,
 $\breve{H}\left (\B^-_{2n_-}\psi(x;E)\right)=
 (E-2\Delta)\B^-_{2n_-}\psi(x;E)$.

In terms of the operators $\A_{n_+}^\pm$
and $\B_{2n_-}^\pm$
we construct the three pairs  of the 
basic ladder operators 
\be\label{A+-ndef}
\mathcal{A}^\pm=\A_{n_+}^- a^\pm \A_{n+}^+\,,
\ee
\be\label{B+-2n-def}
\mathcal{B}^\pm=\B_{2n_-}^- a^\pm \B_{2n-}^+\,,
\ee
 and 
\be\label{C+-2n-n+def}
\mathcal{C}^-=\B_{2n_-}^-\A^+_{n_+}\,,\qquad
\mathcal{C}^+=\A^-_{n_+}\B_{2n_-}^+\,.
\ee
The operators $\mathcal{A}^\pm$ and $\mathcal{B}^\pm$ are
differential operators of  
orders $2n_++1$ and $4n_-+1$, respectively, while
the ladder operators $\mathcal{C}^\pm$ 
are  differential operators of  order 
$n_++2n_-$.
These basic ladder operators  satisfy the relations 
\be\label{HcommuteA,B,C}
[\breve{H},\mathcal{A}^\pm]=
\pm 2 \mathcal{A}^\pm\,,\qquad 
[\breve{H},\mathcal{B}^\pm]=
\pm 2 \mathcal{B}^\pm\,,\qquad 
[\breve{H},\mathcal{C}^\pm]=
\pm  2\Delta\,\mathcal{C}^\pm\,.
\ee
We also have the operator identities 
\be\label{AB2list1}
\mathcal{A}^+\mathcal{A}^-=\mathcal{P}_{\mathcal{A}}(\breve{H})\,,\qquad
\mathcal{A}^-\mathcal{A}^+=\mathcal{P}_{\mathcal{A}}(\breve{H}+2)\,,
\ee
\be\label{AB2list1+B}
\mathcal{B}^+\mathcal{B}^-=\mathcal{P}_{\mathcal{B}}(\breve{H})\,,\qquad
\mathcal{B}^-\mathcal{B}^+=\mathcal{P}_{\mathcal{B}}(\breve{H}+2)\,,
\ee
and
\be\label{AB2list2}
\mathcal{C}^+\mathcal{C}^-=\mathcal{P}_\mathcal{C}(\breve{H})\,,\qquad
\mathcal{C}^-\mathcal{C}^+=\mathcal{P}_\mathcal{C}(\breve{H}+2\Delta)\,,
\ee
where 
\be
\mathcal{P}_{\mathcal{A}}(\breve{H})=
(\breve{H}-2\Delta)
\mathcal{P}_\A(\breve{H}-2)\mathcal{P}_\A(\breve{H})\,,\quad
\mathcal{P}_{\mathcal{B}}(\breve{H})=
\breve{H}
\mathcal{P}_\B(\breve{H}+2\Delta-2)\mathcal{P}_\B(\breve{H}+2\Delta)\,,
\ee
\be
\mathcal{P}_\mathcal{C}(\breve{H})=\mathcal{P}_\B(\breve{H})
\mathcal{P}_\A(\breve{H})\,.
\ee
The polynomial $\mathcal{P}_\A$ of order $n_+$ is  defined here by
$\mathcal{P}_\A(\breve{H})\equiv
\A^-_{n_+}\A^+_{n_+}=\breve{H}\prod_{i=1}^{n_+-1}(\breve{H}-E_i)$,
where $E_i$, $i=1,\ldots ,n_+-1$ , are nonzero
eigenvalues of the 
corresponding excited separated (gapped) physical eigenstates 
$\Psi_i$ of $\breve{H}$. 
Together with zero energy $E_0=0$ of the ground state $\Psi_0$,
the energy values of the $n_+$ gapped physical 
 eigenstates of $\breve{H}$ are the shifted by the  constant 
$2\Delta=2(2n_-+n_+)$ energies of the corresponding 
non-physical eigenstates
$\psi^-_{j_s}$  of $H$ which appear as the seed
states in the minimal  $(\alpha_{n_+})$-scheme.
The permuted product of the intertwining 
operators gives here
  $\A^+_{n_+}\A^-_{n_+}=\mathcal{P}_\A(H)$.
The polynomial 
$\mathcal{P}_\B$ of order $2n_-$ is  defined 
via the relation
$\mathcal{P}_\B(H)\equiv
\B^+_{2n_-}\B^-_{2n_-}=\prod_{j=1}^{2n_-}(H-E_j^-)$,
where  by $E^-_j$ 
we denote  the shifted for the same  constant 
$2\Delta$
energies
of the physical eigenstates $\psi_{i_s}$   of $H$
which are present as the seed states  in the
minimal $(\beta_{2n_-})$-scheme.
For the permuted product of the 
intertwining operators 
we have  $\B^-_{2n_-}\B^+_{2n_-}=
\mathcal{P}_\B(\breve{H}+2\Delta)$.
Let us also note  here a useful relation 
\be\label{complementident}
\mathcal{P}_\A(\breve{H})
\mathcal{P}_\B(\breve{H}+2\Delta)=
\prod_{j=0}^{2n_-+n_+-1}(\breve{H}-2j)\,,
\ee
that reflects  the complementarity of the minimal 
$(\alpha_{n_+})$- and ($\beta_{2n_-}$)-schemes.

The operator $\mathcal{A}^-$  annihilates
all the $n_+$ physical eigenstates 
$\Psi_0,\ldots, \Psi_{n_+-1}$
of the  system $\breve{H}$
whose energies  lie below the infinite equidistant
part of the spectrum and are separated from 
it by some gap of $2n_0$ missing energy levels,
$n_0\geq 1$.
Between these $n_+$ 
separated energy levels  there can appear $g$,
$0\leq g<n_+$,
`internal' gaps each one containing   an
even number of missing energy levels.
We name the $g+1$ sets of energy levels in the lower
separated part of the spectrum which do not contain 
internal gaps as valence bands.
If $g>0$, we denote by $2n_1,\ldots, 2n_g$
the number of missing energy levels in the corresponding 
internal energy gaps assuming that the highest
 value $g$ of index $i$ 
in $n_i$ corresponds here 
to the lowest energy gap in the spectrum.
The total number of the missing energy levels 
$2(n_0+\ldots+n_g)$ is equal to the number
$2n_-$
of the physical eigenstates $\psi_{i_s}$ which participate 
as the seed states in the minimal $(\beta_{2n_-})$-scheme.
In addition to the gapped physical eigenstates 
$\Psi_0$,\ldots,
$\Psi_{n_+-1}$, 
the operator $\mathcal{A}^-$ also annihilates 
the lowest state 
$\Psi_{n_+}$ of energy $E_{n_+}=2\Delta$
in 
the infinite equidistant part
of the spectrum due to the presence of the operator
$a^-$ in its  structure.
Besides, $\mathcal{A}^-$ annihilates 
some $n_+$ non-physical eigenstates of $\breve{H}$.
Note in particular that  
\be
\Psi_0=\A^-_{n_+}\widetilde{\psi^-_{j_+}}=
\B^-_{2n_-}\psi_0\,,\qquad
\Psi_{n_+-1}=\A^-\widetilde{\psi^-_{2n_0}}=
\B^-_{2n_-}\psi_{j_+-2n_0}\,,
\ee
\be
\Psi_{n_+}=\A^-_{n_+}\psi_0=
\B^-_{2n_-}\psi_{2n_-+n_+}\,.
\ee
Here $j_+=2n_-+n_+-1$ is the maximal value
of the index $j_s$ of the non-physical eigenstates
$\psi^-_{j_s}\in \ker\, (\A^-_{n_+})$
from the minimal $(\alpha_{n_+})$-scheme;
it coincides with the maximal value 
of the  index $i_s$
of the physical eigenstates
$\psi_{i_s}\in\ker\,(\B^-_{2n_-})$ from 
the minimal scheme $(\beta_{2n_-})$.
The energy values of the indicated $2n_++1$ physical 
and non-physical eigenstates 
from the kernel of $\mathcal{A}^-$ are 
the roots of the polynomial $\mathcal{P}_{\mathcal{A}}(\breve{H})$
which appears in the first identity in (\ref{AB2list1}).
The kernel of $\mathcal{A}^+$ is spanned by 
the  $n_+$ lowest separated  physical eigenstates 
$\Psi_0$,\ldots, $\Psi_{n_+-1}$,
and  by some $n_++1$ non-physical eigenstates
of $\breve{H}$. The energy values of these
eigenstates from $\ker\,(\mathcal{A}^+)$
correspond to the roots of the polynomial $\mathcal{P}_{\mathcal{A}}(\breve{H}+2)$
which appears in the second
relation  in (\ref{AB2list1}).

The kernel of the ladder operator 
$\mathcal{C}^-$ is 
spanned by $n_++2n_-$ physical states,
$n_+$ of which, $\Psi_0$,\ldots, $\Psi_{n_+-1}$,  
correspond to the lowest separated 
(gapped) energy values. The  other  $2n_-$ eigenstates of $\breve{H}$ 
in the kernel of  $\mathcal{C}^-$ are 
the physical states $\A^-_{n_+}\psi_{i_s}$
in a lower part
of the infinite equidistant part of the spectrum,
where $\psi_{i_s} \in \ker\,(\B^-_{2n_-})$.
 The number of those `supplementary' states in the lower part
of the equidistant spectrum which are not annihilated 
by $\mathcal{C}^-$ and whose eigenvalues
lie below  the highest energy value 
of a physical eigenstate from $\ker\, (\mathcal{C}^-)$ 
is equal to $n_+$.
The number of the sequential  lowest states 
at the very bottom of the  equidistant 
infinite part of the spectrum of $\breve{H}$ 
which are not annihilated by $\mathcal{C}^-$ 
is equal to the number of the 
physical states 
in the lowest valence band of the separated part of the spectrum. 
The kernel of $\mathcal{C}^+$ will be spanned
by some $n_++2n_-$ non-physical eigenstates.
The energies of the corresponding eigenstates from the 
kernels of the ladder operators $\mathcal{C}^-$ and 
$\mathcal{C}^+$ correspond to the roots of the polynomials 
in $\breve{H}$ which appear in the first and the second identities in
(\ref{AB2list2}). 

The kernel of the ladder operator $\mathcal{B}^-$ 
contains $g+1$ physical eigenstates, each one lying
 at the very  bottom of each valence band. Besides, 
it also contains the physical eigenstate $\Psi_{n_+}$
of energy $E_{n_+}=2\Delta$ which is the lowest state
of the equidistant infinite part of the spectrum.
In addition, $\ker\, (\mathcal{B}^-)$ 
contains $2n_-$ non-physical eigenstates 
of the form $\Psi^{non}_{i_s}\equiv \B^-\widetilde{\psi_{i_s}}$,
which correspond to the missing energy values 
in the gaps. Finally, it also involves $2n_- - (g+1)$ 
non-physical eigenstates
of the form $\widetilde{\Psi^{non}_{i_s}}$
for all values of the index $i_s$
except those $g+1$ values,  each one of which 
corresponds to a lowest eigenstate
in each gap.  
The kernel of the increasing 
ladder operator $\mathcal{B}^+$ 
contains $g+1$ physical eigenstates
whose eigenvalues lie at the top of
each valence band. It also involves  $4n_- - g$ 
non-physical eigenstates from the gaps.
Eigenvalues of the eigenstates  from the kernels
of the ladder operators $\mathcal{B}^-$ and 
$\mathcal{B}^+$ correspond to the roots of the 
polynomials which appear, respectively,  in the first and 
the second identities
in (\ref{AB2list1+B}).

 The relations of the form (\ref{ABa3}) and 
(\ref{B2A2a})  are valid
for the basic intertwining operators 
of the arbitrary REQHO system
we consider here.
Indeed, the operator $\B^-_{2n_-}$ annihilates all the $2n_-$
physical eigenstates  $\psi_{i_s}$ of the QHO which
participate as the seed states in the $(\beta_{2n_-})$-scheme.
On the other hand, when $\B^-_{2n_-}$  acts on the $n_+$  
`supplementary'  eigenstates  in the lower part of the 
spectrum of the QHO, it transforms these states
into the separated  lowest $n_+$ 
physical eigenstates 
of the REQHO system
$\breve{H}$ which constitute the kernel 
of the intertwining operator $\A_{n_+}^+$. 
In particular, as we saw $\B^-_{2n_-}$ maps the ground-state $\psi_0$ 
of the shifted QHO, $H\psi_0=2\Delta\psi_0$, 
into the zero energy ground-state $\Psi_0$ of $\breve{H}$,
$\B^-_{2n_-}\psi_0=\Psi_0$, 
$\breve{H}\Psi_0=0$.
We conclude then that the composite operator 
$\A_{n_+}^+\B^-_{2n_-}$ annihilates all the $2n_-+n_+$ 
eigenstates $\psi_n$
of the QHO with $n=0,\ldots,2n_-+n_+ -1$.  
But the same job is made by the lowering 
ladder operator 
$(a^-)^{2n_-+n_+}$ of the QHO.
{}From here we obtain the operator equalities
 \be\label{AB-n+2n-}
 \A^+_{n_+}\B^-_{2n_-}=(-1)^{n_+}(a^-)^{2n_-+n_+}\,,\qquad
 \B^+_{2n_-} \A^-_{n_+}=(-1)^{n_+}(a^+)^{2n_-+n_+}\,.
 \ee
 The  relations (\ref{AB-n+2n-}) reflect
 the complementary nature of the 
 involved minimal $(\alpha_{n_+})$- and $(\beta_{2n_-})$- schemes.
The identities (\ref{AB-n+2n-}) are employed
to establish the operator identities in (\ref{AB2list2}).
They are also essential for  the analysis of the
kernels of the basic ladder operators.

Similarly to the simplest case of the REQHO system,
one can construct other pairs  of secondary   
ladder operators  different from the described 
basic  ladder operators  
$\mathcal{A}^\pm$,  $\mathcal{B}^\pm$ and $\mathcal{C}^\pm$.
This can be done effectively by introducing 
additional factors  $(a^\pm)^n$
inside the structure of these operators\,:
$\mathcal{A}^\pm_n\equiv \A^-_{n_+}(a^\pm)^n\A^+_{n_+}$, 
$\mathcal{B}^\pm_n\equiv \B^-_{2n_-}(a^\pm)^n\B^+_{2n_-}$,
$n=1,\dots$,
$\mathcal{A}^\pm_1=\mathcal{A}^\pm$,
$\mathcal{B}^\pm_1=\mathcal{B}^\pm$,
and
$\mathcal{C}^+_{n+1}\equiv \A^-_{n_+}(a^+)^n\B_{2n_-}^+$,
$\mathcal{C}^-_{n+1}\equiv \B_{2n_-}^- (a^-)^n \A^+_{n_+}$, 
where $n=0,\ldots$,
$\mathcal{C}^\pm_1=\mathcal{C}^\pm$.
 One can also consider
the operators $\mathcal{C}^-_{-n}=\B^-_{2n_-}(a^+)^n\A^+_{n_+}$,
$\mathcal{C}^+_{-n}=\A^-_{n_+} (a^-)^n \B^+_{2n_-}$ with
$n=1,\ldots,2n_-+n_+-1$. In 
$\mathcal{C}^\pm_{-n}$ we restrict the values of  
the  index $n$ from above
having in mind the identity  
\be
\mathcal{C}^-_{-(2n_-+n_+)}=
(-1)^{n_+}\,\mathcal{P}_\A(\breve{H})\,\mathcal{P}_{\B}(\breve{H}+2\Delta)\,,
\ee
see Eq. (\ref{complementident}), 
and so, for $n>2n_-+n_+-1$ these operators 
do not provide  essentially  new structures. 
The secondary, higher-order ladder operators 
can also be obtained 
by taking  the compositions
 of the intertwining operators of the 
corresponding 
$(\alpha)$-, $(\beta)$- and  $(\gamma)$-schemes.
They also are generated via the 
composition  of the basic ladder operators
$\mathcal{A}^\pm$, $\mathcal{B}^\pm$ 
 and $\mathcal{C}^\pm$.
In particular,  the quadratic compositions 
of $\mathcal{A}^\pm$
 and $\mathcal{C}^\pm$
are given by (\ref{AB2list1}), (\ref{AB2list2}),
and by the relations 
\be\label{AB2list1+}
(\mathcal{A}^+)^2=\mathcal{P}_\A(\breve{H}-2) \mathcal{A}^+_2\,,
\quad
(\mathcal{C}^+)^2=(-1)^{n_+}\mathcal{C}^+_{2n_-+n_++1}\,,\quad
\mathcal{A}^+\mathcal{C}^-=(-1)^{n_+}(\breve{H}-2\Delta)\mathcal{A}^-_{2n_-+n_+-1}\,,
\ee
\be\label{AB2list2+}
\mathcal{A}^+\mathcal{C}^+=\mathcal{P}_\A(\breve{H}-2) \mathcal{C}^+_1\,,\quad
\mathcal{A}^-\mathcal{C}^-=(-1)^{n_+} \mathcal{A}^-_{2n_-+n_++1}\,,\quad
\mathcal{A}^-\mathcal{C}^+=\mathcal{P}_\A(\breve{H}+2)\cdot \mathcal{C}^+_{-1}\,,
\ee
and by the relations conjugate to (\ref{AB2list1+}) and (\ref{AB2list2+}).
The relation 
\be\label{AC-relation}
(\mathcal{A}^-)^{2n_-+n_+}=
(-1)^{n_+}\prod_{l=0}^{2n_-+n_+-1}\mathcal{P}_\A(\breve{H}+2l)\cdot
\mathcal{C}^-
\ee
shows that as in the considered particular cases 
of the REQHO systems, 
the ladder operators $\mathcal{C}^\pm$ can be
generated  by the operators $\mathcal{A}^\pm$.
Also,  the following operator identity is valid: 
\be
(\mathcal{A}^-)^{2n_-+n_+-1}=
(-1)^{n_+}\frac{1}{\breve{H}}\prod_{j=0}^{2n_-+n_+-2}
\mathcal{P}_\A(\breve{H}+2j)\, \mathcal{C}^-_{-1}\,.
\ee 
Here the operator multiplier before $\mathcal{C}^-_{-1}$
is  the polynomial of order $n_+(2n_-+n_+-1)-1$
in $\breve{H}$
since the $j=0$ term $\mathcal{P}_\A(\breve{H})$ 
in the product is equal to 
the factor $\breve{H}$ which cancels the multiplier 
$\frac{1}{\breve{H}}$ before 
the product symbol.
We also have the  identity 
which relates the operators $\mathcal{A}^-$ and 
$\mathcal{B}^-$,
\be\label{B-indenityA-}
(\breve{H}-2\Delta+2)\mathcal{P}_\A(\breve{H}+2)\,\mathcal{B}^-=
(\breve{H}+2)\mathcal{P}_\B(\breve{H}+2\Delta)\,\mathcal{A}^-\,.
\ee
The analogous identity for $\mathcal{A}^+$ and 
$\mathcal{B}^+$ is obtained from 
(\ref{B-indenityA-}) by Hermitian conjugation.
In particular cases of the three REQHO systems considered in the previous
two sections, relation (\ref{B-indenityA-})
reduces to the first relation in (\ref{BnACH-4}) and to
the
identities (\ref{B-A-H2H4H6}) and 
(\ref{BAn+=2n-=1}). 

In conclusion of this section, let us  show that the trinity 
$(\mathcal{A}^\pm,\mathcal{B}^\pm,\mathcal{C}^\pm)$
of the pairs of the lowering  and raising ladder operators 
allows us to generate an arbitrary physical eigenstate from the 
ground state $\Psi_0$, and as a consequence, any two 
physical eigenstates can be related by the appropriate consecutive 
action of the basic ladder operators from the trinity.
First, from the described properties of the operators 
and commutation relations (\ref{HcommuteA,B,C}) 
it follows that in the equidistant infinite part of the spectrum 
any two eigenstates can be related by
the ladder operators $\mathcal{A}^\pm$ and $\mathcal{B}^\pm$ 
in the same way as  the ladder operators $a^\pm$ relate the states in the 
QHO system. 
The only difference will appear in the numerical 
coefficients which have to be included into the composition of the
indicated basic operators when we work with the normalized
eigenstates. 
If a valence band contains more than one eigenstate,
different states in this band can be connected by application to them of the
appropriate degrees of the lowering and raising operators   $\mathcal{B}^-$ and 
$\mathcal{B}^+$. Note that within the valence band  with $n_i$ states 
these operators satisfy the identity $(\mathcal{B}^\pm)^{n_i}=0$.
Recall also that the lowest state $\Psi_{n_+}$ in the equidistant 
infinite part of the spectrum is related with the ground state 
by the action of the ladder operators $\mathcal{C}^\pm $:
 $\Psi_0=\mathcal{C}^-\Psi_{n_+}$ and 
$\Psi_{n_+}=\mathcal{C}^+\Psi_0$.
In the same way 
one can relate any  state $\Psi_n$ 
of energy $0<E_n<2\Delta$ with $0<n\leq {n_+-1}$ from 
the separated part of the spectrum with the corresponding
state $\Psi_{n_++n}$ of energy $E_n+2\Delta$ from the equidistant 
infinite part of the spectrum. Then, if a REQHO system contains more than 
one valence band, the ground state $\Psi_0$ from the 
lowest valence band can be related to some state $\Psi_l$ with eigenvalue 
$E_l$ from some higher valence band, for instance,
by the following composition of the ladder operators:
$\Psi_l=\mathcal{C}^-(\mathcal{A}^+)^{r_l}\mathcal{C}^+\Psi_0$,
$\Psi_0=\mathcal{C}^- (\mathcal{A}^-)^{r_l}\mathcal{C}^+\Psi_l$, 
where $r_l=E_l/2$. This shows finally  
that the ladder operators from the  trinity are the 
spectrum-generating operators for  the REQHO system of a 
general form. 

\begin{figure}[htbp]
 \begin{center}
\includegraphics[scale=0.7]{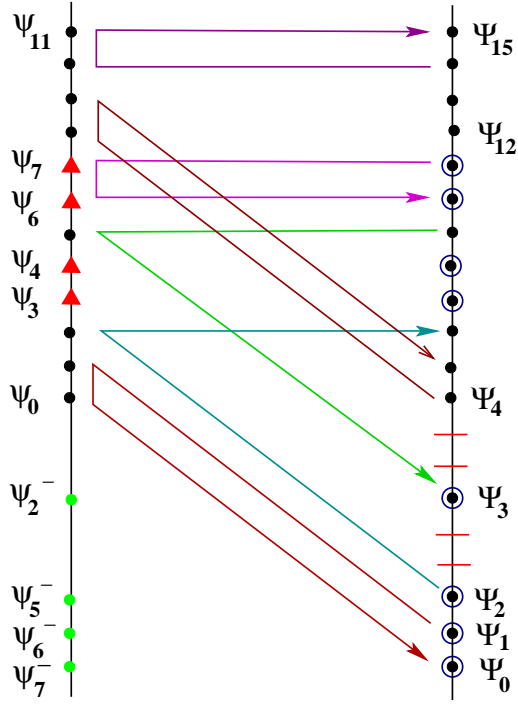}
\caption{Example of the REQHO system with two gaps and four separated states organized into
 two valence 
bands of one and three states. The  system is generated by 
the DCKATs based on the complementary 
schemes $(\alpha_4)=\{\psi^-_2,\psi^-_5,\psi^-_6,\psi^-_7\}$
and $(\beta_4)=\{\psi_3,\psi_4,\psi_6,\psi_7 \}$. 
The following action 
of the ladder operators is shown: $\Psi_{15}=\mathcal{A}^+\Psi_{14}$,
$\Psi_{10}=\mathcal{A}^-\Psi_{11}$, $\Psi_{5}=\mathcal{B}^+\Psi_{4}$,
$\Psi_{0}=\mathcal{B}^-\Psi_{1}$,
$\Psi_{3}=\mathcal{C}^-\Psi_{9}$, $\Psi_{6}=\mathcal{C}^+\Psi_{2}$,
where $\mathcal{A}^\pm=\A^-_4 a^\pm\A^+_4$,
$\mathcal{B}^\pm=\B^-_4 a^\pm\B^+_4$, $\mathcal{C}^-=\B^-_4\A^+_4$,
$\mathcal{C}^+=\A^-_4\B^+_4$. On the right,  eight encircled states
belong to the kernel of $\mathcal{C}^-$ and their energies correspond to 
the energies of the seed states from the complementary
$(\alpha_4)$- and $(\beta_4)$-  schemes.
If  the equidistant part of the 
spectrum of the REQHO system
is moved down for the distance $2\Delta=2(n_+ + 2n_-)=16$, that corresponds 
to the difference between energies of the lowest state $\Psi_4$
in the equidistant
part of the spectrum 
and the ground state $\Psi_0$, the encircled energy levels
in the equidistant part of the spectrum 
take exactly a position of missing energy levels in 
the low, separated  part of the spectrum.
This refelects a complementarity of the minimal 
$(\alpha_4)$- and $(\beta_4)$-  schemes, according to which 
we  have  $\Psi_0=\A^-_4\widetilde{\psi^-_7}=\B^-_4\psi_0$,
$\Psi_1=\A^-_4\widetilde{\psi^-_6}=\B^-_4\psi_1$,
$\Psi_2=\A^-_4\widetilde{\psi^-_5}=\B^-_4\psi_2$,
$\Psi_3=\A^-_4\widetilde{\psi^-_2}=\B^-_4\psi_5$,
$\Psi_{4+n}=\A^-_4\psi_n=\B^-_4\psi_{8+n}$, 
$n=0,1,\ldots.$}
\label{Fig1}
\end{center}
\end{figure}

The described properties of the REQHO systems 
of a general form are illustrated by  Figure~\ref{Fig1}.

\vskip1cm

\section{Summary and outlook}\label{secSummary}

In conclusion, we summarize the obtained results
and indicate some interesting problems  for 
further investigation.

The  REQHO  system of a general form is characterized
by $n_+\geq 1$   low-lying  energy levels 
which are separated from 
the higher  equidistant infinite part of the spectrum by some 
gap  of  an even number $2n_0\geq 2$ of missing  levels. 
Between  separated energy levels there can be 
additional gaps of an  even number of missing levels 
in each such a gap. As a result,
in the lower  part
of the spectrum there appears a total number $2n_-\geq 2n_0$
of missing levels, and  the separated part of the  spectrum is organized  
into $(g+1)\geq 1$  `valence bands'.
Such a peculiar structure of the spectrum of the 
REQHO system characterized by the three integer numbers
$(n_+,2n_-,g+1)$ is detected and reflected 
by the trinity $(\mathcal{A}^\pm,\mathcal{B}^\pm,\mathcal{C}^\pm)$
of the pairs of the lowering and raising ladder operators.
Any raising or lowering  ladder operator from the trinity when acts on a 
physical eigenstate either transforms it into another physical 
eigenstate with the changed value of energy, or annihilates it.

The separated states 
are detected by 
$\mathcal{A}^-$ and $\mathcal{A}^+$ that are differential
operators of order $2n_++1$.  Each one of these two
operators annihilates all the $n_+$ states
$\Psi_0$,\ldots, $\Psi_{n_+-1}$
 in the valence bands.
The operator $\mathcal{A}^-$
annihilates in addition the lowest state 
$\Psi_{n_+}$
in the equidistant infinite part of the spectrum.
In this aspect the lowering ladder operator $\mathcal{A}^-$
has  properties very similar to those
 of the Lax-Novikov integral in reflectionless
systems we discussed in Section~\ref{secIntro}. 
Besides, the operators $\mathcal{A}^-$
and $\mathcal{A}^+$ annihilate some 
$n_+$ and $n_++1$ non-physical eigenstates, respectively.
Due to the relation
$[\breve{H},\mathcal{A}^\pm]=\pm 2\mathcal{A}^\pm$,
the $\mathcal{A}^\pm$
act in the equidistant infinite  part of the spectrum
as the spectrum-generating operators 
like the ladder operators $a^\pm$ in the 
QHO system. Namely, 
they transform  physical eigenstates 
$\Psi_j$ with $j\geq n_+$
into the states $\Psi_{j\pm 1}$ 
by shifting the energy values in 
$\pm 2$, and $\mathcal{A}^-\Psi_{n_+}=0$.
It is also worth to note here that 
the quadratic in the
ladder operators $\mathcal{A}^\pm$
relations in (\ref{AB2list1}) 
are analogous to the 
Burchnall-Chaundy polynomial 
identity
\cite{LaxNov,exosusy,BurCha}
that relates the Lax-Novikov  integral 
with the corresponding Hamiltonian 
of a reflectionless (or a finite-gap) system,
and underlies the modern theory of 
integrable systems \cite{Krich}.

 The ladder operators $\mathcal{B}^\pm$
 are differential operators of order $4n_-+1$,
 and each of them  effectively counts
 the number $g+1$  of the valence bands
 and measures
  the size of each valence band.
 This is done as follows. The lowering operator
 $\mathcal{B}^-$  annihilates each physical eigenstate
 which lies at the very bottom of each valence band,
 whereas the raising operator  $\mathcal{B}^+$ 
annihilates each state  at the very top of
each valence band.
So, if a valence band contains only one state,
this state 
is annihilated by both $\mathcal{B}^-$ and $\mathcal{B}^+$,
and in such a one-dimensional 
valence band the action of $\mathcal{B}^\pm$
is similar to that of $\mathcal{A}^\pm$.
However, if we have a valence band with more than one 
state, 
the operators $\mathcal{B}^+$ and $\mathcal{B}^-$,
unlike $\mathcal{A}^\pm$,
act in such a band as the raising 
and lowering operators,
$[\breve{H},\mathcal{B}^\pm]=\pm 2\mathcal{B}^\pm$,
which satisfy there the 
relations $(\mathcal{B}^\pm)^{n_i}=0$,
where $n_i$ is the number of 
states in the band.
Like $\mathcal{A}^-$, the  operator $\mathcal{B}^-$ 
annihilates the lowest state $\Psi_{n_+}$ 
in the equidistant infinite
part of the spectrum.
The kernels of 
$\mathcal{B}^-$  and $\mathcal{B}^+$ 
also include some $4n_--g-1$
and  $4n_--g$ non-physical eigenstates, respectively.
Similarly to $\mathcal{A}^\pm$, 
in the equidistant infinite part of the spectrum
the operators $\mathcal{B}^\pm$ also act 
as the spectrum-generating operators.

Under the action of the ladder operators $\mathcal{A}^\pm$ 
and $\mathcal{B}^\pm$ the $n_+$ states 
from the valence bands turn out to be completely disconnected 
from the physical states
in the equidistant infinite part of the spectrum.
Both parts of the spectrum are connected
by means of the third pair of the 
mutually conjugate ladder operators $\mathcal{C}^+$ and 
$\mathcal{C}^-$, that are 
 differential 
operators of order $n_++2n_-$. 
The kernel of the lowering operator 
$\mathcal{C}^-$ is spanned by physical
eigenstates, $n_+$ of which 
correspond to  all the $n_+$ eigenstates 
from the valence bands.
The rest $2n_-$ states from  
$\ker\,(\mathcal{C}^-)$ 
are some eigenstates 
with energy levels lying in the low part
of the  equidistant  infinite part of the spectrum.
The positions of those $2n_-$ energy levels 
correspond to the missing energy levels 
in the gaps moved up for the distance 
$2\Delta=2n_++4n_-\geq 6$ which is 
exactly equal   to  the distance  between  the energy levels of the 
ground state $\Psi_0$ and the lowest state $\Psi_{n_+}$ 
in the equidistant infinite part  of the spectrum.
The kernel of $\mathcal{C}^+$ is spanned by
some non-physical eigenstates only.
Due to the relation 
$[\breve{H},\mathcal{C}^\pm]=\pm 2\Delta\, \mathcal{C}^\pm$,
the operators $\mathcal{C}^+$ and $\mathcal{C}^-$
act as the 
raising and lowering operators
changing the energy for $\pm 2\Delta$.
All the states from the valence bands
are obtained by application of the lowering 
operator $\mathcal{C}^-$ to those low-lying  states with energies 
$2\Delta\leq E<4\Delta$ in the 
equidistant infinite part of the spectrum 
which are not annihilated by it.
In particular, the lowest state 
$\Psi_{n_+}$ with energy $E_{n_+}=2\Delta$ 
in the equidistant infinite part of the spectrum is transformed by $\mathcal{C}^-$
into the ground-state $\Psi_0$ of zero energy.
The state $\Psi_{n_+}$, in turn,  
can be obtained from $\Psi_0$ 
by action of the raising operator $\mathcal{C}^+$, and 
is also generated 
from the state $\Psi_{n_++\Delta}$
of energy $E_{{n_++\Delta}}=4\Delta$ by applying to the 
latter the lowering operator   $\mathcal{C}^-$.
As a consequence of the described properties, 
any two states in the spectrum of  a REQHO system can be
related by an appropriate consecutive 
action of the basic ladder operators from the trinity.
In particular, arbitrary excited state from any valence band
or from the equidistant infinite part of the spectrum 
can be obtained from the ground state $\Psi_0$.
This means that the basic ladder operators from the trinity 
are the spectrum-generating operators of the REQHO system.

The energies of all the physical and non-physical eigenstates
of the kernels of the lowering operators 
$\mathcal{A}^-$, $\mathcal{B}^-$ and $\mathcal{C}^-$
are   the roots of the corresponding polynomials in $\breve{H}$  
which appear on the right hand side in  the first relations from equations 
(\ref{AB2list1}), (\ref{AB2list1+B}) and (\ref{AB2list2}), respectively.
 The eigenvalues of the physical and non-physical eigenstates 
 from the kernels of the conjugate operators
 $\mathcal{A}^+$, $\mathcal{B}^+$ and $\mathcal{C}^+$
are the roots of the corresponding polynomials 
which appear in the second operator identity relations
in the same equations. 
The   basic ladder operators $\mathcal{A}^-$ and $\mathcal{B}^-$ 
satisfy the  two-term  identity relation (\ref{B-indenityA-})  
which is  linear in 
both of these operators but involve the 
coefficients that 
are certain polynomials in the Hamiltonian $\breve{H}$.
The presence of such polynomial coefficients reflects 
a difference in action of these operators on the states
in a separated part of the spectrum. 
The operators $\mathcal{A}^-$ and $\mathcal{C}^-$
are related by the operator identity of the form
(\ref{AC-relation}). 
Proceeding from these relations,
one can obtain the identity  that  
relates the operators $\mathcal{B}^-$ 
and $\mathcal{C}^-$, and by conjugation 
one can find the identities that relate 
the raising operators of the trinity.

The operators $\mathcal{A}^\pm$ 
are constructed as the ladder operators $a^\pm$ of the 
QHO dressed by means of the Darboux-Crum-Krein-Adler  intertwining 
operators $\A^-_{n_+}$ and $\A^+_{n_+}$ constructed on the basis of 
the minimal set of $n_+$ non-physical eigenstates
of the QHO which are used as the seed states
in the corresponding DCKAT based on the $(\alpha_{n_+})$-scheme,
see Eq. (\ref{A+-ndef}). 
The operators $\mathcal{B}^\pm$
are constructed in the same way with the help of the intertwining operators
$\B^-_{2n_-}$ and $\B^+_{2n_-}$ obtained 
on the basis 
of the minimal set of the $2n_-$ physical eigenstates 
which are employed as the seed states in the DCKAT 
in the corresponding $(\beta_{2n_-})$-scheme,
see Eq. (\ref{B+-2n-def}). 
The operators $\mathcal{C}^\pm$ can be obtained 
as the composition (\ref{C+-2n-n+def}) of the corresponding 
intertwining operators from both indicated 
schemes of the DCKATs. 
The minimal schemes  $(\alpha_{n_+})$ and 
$(\beta_{2n_-})$ are complementary, 
what  is reflected in particular by the relations 
(\ref{complementident}) and
(\ref{AB-n+2n-}).
The  secondary, higher-order ladder operators can be constructed 
in analogous way by dressing the higher-order 
ladder operators $(a^\pm)^n$ of the QHO, or
by the composition of the appropriate 
intertwining operators from the non-minimal $(\alpha_{n_++n})$
and $(\beta_{2n_-+n})$ schemes,
or by employing the intertwining operators from the corresponding 
intermediate ($\gamma$)-type schemes 
which use  both physical and non-physical eigenstates
of the QHO as the seeds states of the corresponding 
DCKATs. The  secondary ladder operators can  also be generated
via the appropriate composition of the 
basic (primary)  ladder operators $\mathcal{A}^\pm$,
 $\mathcal{B}^\pm$ and $\mathcal{C}^\pm$.
 
 It seems to be interesting to investigate the 
 quantum mechanical systems related to the
 exceptional Laguerre and Jacobi orthogonal polynomials
 in the light of the results on the ladder operators
 obtained here. 
 The  results of such an investigation will be presented elsewhere.

 The ladder operators $\mathcal{B}^\pm$ 
 have a nature of the polynomially deformed 
  bosonic creation and annihilation
  operators in the equidistant infinite part 
  of the spectrum.  On the other hand, 
  these operators act trivially on
 the one-state valence bands 
whose singleton states  are annihilated by both 
the lowering $\mathcal{B}^-$
and the raising $\mathcal{B}^+$ 
operators. The same operators
reveal the properties of 
the deformed fermionic
creation and annihilation operators
in the 
valence bands consisting from 
two states. 
They have the properties of the deformed para-fermion
creation and annihilation operators of order $n>2$ in those 
valence bands which contain   $n>2$
eigenstates of $\breve{H}$.   
The interesting question is then if there exist some 
concrete physical systems which would 
reveal  the spectrum of the 
REQHO systems. 
If so, it seems that the trinity of the ladder operators 
should play a fundamental  role in the physics 
associated  with such  systems. 
In the same direction the interesting question is whether
the quantum mechanical REQHO systems and the    
structures associated with them 
can be generalized somehow for the case of  
the quantum fields.  

In \cite{MarQue1+}, the family of  the REQHO 
systems with two separated states 
generated by  non-physical seed states
$\psi^-_{m_1}$ and $\psi^-_{m_2}$, 
$m_2-m_1\equiv\ell=1+2r$, $r=0,1,\ldots$,
$m_1=2k$, $k=1,\ldots$,  
was considered.
For such class of the systems,
there the  lowering, $c$,  and 
increasing, $c^\dagger$, ladder operators of the differential order
$2+\ell$ were constructed
by employing auxiliary systems  some of which 
are singular and have a nature similar
to that of the isotonic oscillator (\ref{Hisodef}).
In the simplest case $m_1=2$ and $m_2=3$
such a system corresponds to the REQHO system (\ref{potentialx4})
we considered in Section \ref{sec+2systems}.
Like our fifth order ladder operator $\mathcal{B}^-$,
the third order ladder operator $c$ from \cite{MarQue1+} 
annihilates the
ground state $\Psi_0$ and the lowest state $\Psi_2$ 
in the infinite equidistant part of the spectrum.
The increasing operator  $c^\dagger$ like 
our $\mathcal{B}^+$ annihilates the excited state $\Psi_1$
in the separated two-state lower part of the spectrum.
In the systems with $\ell>1$,  however, the kernel of the increasing operator
$c^\dagger$ still includes only one physical state which is, again, 
the separated state $\Psi_1$, while our $\mathcal{B}^+$
operator annihilates both separated states $\Psi_0$ and $\Psi_1$. 
In addition to the separated ground 
state $\Psi_0$ and the lowest state $\Psi_2$ 
in the equidistant part of the spectrum,
the kernel of  the lowering operator $c$ in this case includes
also the $\ell-1$ excited states 
$\Psi_3,\ldots,\Psi_{2+\ell}$ in the equidistant part of the spectrum.
In this aspect, the lowering operator $c$ from  \cite{MarQue1+} 
has some similarity 
with our operator $\mathcal{C}^-$. But our  
ladder operator $\mathcal{C}^-$ is
of differential order $2m_1+2r$ and its kernel includes
some $2m_1+2r-2$ excited states in the equidistant 
part of the spectrum together with
both separated states 
$\Psi_0$ and $\Psi_1$. 
Thus, in the case of $\ell>1$ the nature of the 
operators $c$ and $c^\dagger$  in the sense  
of the physical states which  they annihilate  is different from
the nature of any of
our lowering and increasing ladder operators 
$\mathcal{A}^-$, $\mathcal{A}^+$,
$\mathcal{B}^-$, $\mathcal{B}^+$ and $\mathcal{C}^-$, $\mathcal{C}^+$.
It would be interesting to investigate whether the analogs 
of the ladder operators $c$ and $c^\dagger$ from  \cite{MarQue1+}  
can be 
constructed for REQHO systems containing more than two 
 states in the lower separated part of the spectrum,
 and what is  the exact  relation of such ladder operators with
our trinity $(\mathcal{A}^\pm$,
$\mathcal{B}^\pm$, $\mathcal{C}^\pm$)  of the ladder operators.

\vskip0.2cm

\noindent {\large{\bf Acknowledgements} } 
\vskip0.4cm

JFC and MSP acknowledge support from
research projects FONDECYT 1130017 (Chile),
Proyecto Basal USA1555 (Chile),
MTM2015-64166-C2-1 (MINECO, Madrid) and DGA E24/1 (DGA, Zaragoza).
MSP is grateful  for the warm hospitality at  Zaragoza University.
JFC thanks for the kind hospitality at Universidad de Santiago de Chile.

\vskip2cm



\begin{thebibliography}{99}

\bibitem{KaiMos}
I. Kay and H. E. Moses, 
 \emph{``Reflectionless transmission through dielectrics and scattering potentials,"},
 \href{http://aip.scitation.org/doi/abs/10.1063/1.1722296}{J. Appl. Phys. {\bf 27}, 1503 (1956)}.


\bibitem{Darb}
G. Darboux,\emph{ ``Sur une proposition relative aux \'equations 
lin\'eaires,"}  C. R. Acad. Sci Paris {\bf 94},  1456 (1882).


\bibitem{Schr}
  E.~Schr\"odinger,
  \emph{``A method of determining quantum-mechanical eigenvalues and eigenfunctions,''}
  Proc.\ Roy.\ Irish Acad.\ (Sect.\ A) {\bf 46}, 9  (1940).

\bibitem{InfHull} 
  L.~Infeld and T.~E.~Hull,
\emph{  ``The factorization method,''}
  \href{http://journals.aps.org/rmp/abstract/10.1103/RevModPhys.23.21}{
  Rev.\ Mod.\ Phys.\  {\bf 23}, 21 (1951)}.

\bibitem{Crum}
M. M. Crum, \emph{``Associated Sturm-Liouville systems,"}
\href{http://qjmath.oxfordjournals.org/content/6/1/121.full.pdf+html}{
Quart. J. Math. Oxford  {\bf 6}, 121 (1955)}.

 \bibitem{MatSal}
V. B. Matveev and M. A. Salle, \textsl{Darboux Transformations and Solitons}
(Springer, Berlin,
1991).

\bibitem{CR00} J. F. Cari\~nena and  A. Ramos, 
 \emph{``Riccati equation, factorization method and shape invariance,"}
   \href{http://www.worldscientific.com/doi/abs/10.1142/S0129055X00000502}{
Rev.\  Math.\ Phys.\ {\bf 12}, 1279  (2000)}
\href{http://arxiv.org/abs/math-ph/9910020}{\textcolor{magenta}{[arXiv:math-ph/9910020]}}.

   \bibitem{CR01} J. F. Cari\~nena, D. J. Fern\'andez  and  A. Ramos, 
  \emph{  ``Group theoretical approach to the intertwined  Hamiltonians,''}
 \href{http://www.sciencedirect.com/science/article/pii/S0003491601961792}{
Ann.\  Phys.\  {\bf 292}, 42  (2001)}
\href{http://arxiv.org/abs/math-ph/0311029}{\textcolor{magenta}{[arXiv:math-ph/0311029]}}.

\bibitem{MelRos}
B. Mielnik and O. Rosas-Ortiz,
\emph{  ``Factorization: little or great algorithm?,''}
  \href{http://iopscience.iop.org/article/10.1088/0305-4470/37/43/001/meta}{
J. Phys. A  {\bf 37}, 10007 (2004)}. 
  
\bibitem{CR08} J. F. Cari\~nena and  A. Ramos, 
  \emph{  ``Generalized B\"acklund-Darboux transformations in one-dimensional 
quantum mechanics,''}
\href{http://www.worldscientific.com/doi/abs/10.1142/S0219887808002989}{
Int.\ J. Geom.\ Methods Mod. \ Phys.\ {\bf 5}, 605 (2008)}. 
  
\bibitem{GroNev}
  D.~J.~Gross and A.~Neveu,
  \emph{``Dynamical symmetry breaking in asymptotically free field theories,''}
  \href{http://journals.aps.org/prd/abstract/10.1103/PhysRevD.10.3235}{Phys.\ Rev.\ D {\bf 10}, 3235 (1974)}.

\bibitem{DasHasNev}
R. F.  Dashen, B. Hasslacher and A. Neveu,
  \emph{``Semiclassical bound states in an asymptotically free theory,"}
  \href{http://journals.aps.org/prd/abstract/10.1103/PhysRevD.12.2443}{Phys.\ Rev.\ D {\bf 12}, 2443 (1975)}.


\bibitem{CamBis}
D. K.  Campbell and  A. R. Bishop,
  \emph{``Soliton excitations in polyacetylene and relativistic field theory models,"}
 \href{http://www.sciencedirect.com/science/article/pii/055032138290089X}{Nucl.\ Phys.\ B {\bf 200}, 297 (1982)}.

\bibitem{CoJaPl} 
 F.~Correa, V.~Jakubsky and M.~S.~Plyushchay,
   \emph{ ``Aharonov-Bohm effect on AdS(2) and nonlinear supersymmetry of reflectionless P\"oschl-Teller system,''}
 \href{http://www.sciencedirect.com/science/article/pii/S000349160900044X}{ Annals Phys.\  {\bf 324}, 1078 (2009)}
 \href{https://arxiv.org/abs/0809.2854}{\textcolor{magenta}{ [arXiv:0809.2854 [hep-th]]}}.

\bibitem{LaxNov} 
  A.~Arancibia, J.~Mateos Guilarte and M.~S.~Plyushchay,
 \emph{  ``Effect of scalings and translations on the supersymmetric 
 quantum mechanical structure of soliton systems,''}
  \href{http://journals.aps.org/prd/abstract/10.1103/PhysRevD.87.045009}{
  Phys.\ Rev.\ D {\bf 87},  045009 (2013) }
 \href{http://arxiv.org/abs/1210.3666}{\textcolor{magenta}{  [arXiv:1210.3666 [math-ph]]}}.

\bibitem{BBEIM}
E. D. Belokolos, A. I. Bobenko, V. Z. EnolÕskii, A. R. Its, and V. B. Matveev, 
  \textsl{Algebro-Geometric Approach to Nonlinear Integrable Equations},
(Springer, Berlin, 1994).


\bibitem{GesHol}
 F. Gesztesy and H. Holden,
   \textsl{Soliton Equations and their Algebro-Geometric Solutions}, (Cambridge Univ. Press, 2003). 
   
 \bibitem{exosusy} 
 F.~Correa, V.~Jakubsky, L.~M.~Nieto and M.~S.~Plyushchay,
  \emph{ ``Self-isospectrality, special supersymmetry, and their effect 
  on the band structure,''}
  \href{http://journals.aps.org/prl/abstract/10.1103/PhysRevLett.101.030403}{Phys.\ Rev.\ Lett.\  {\bf 101}, 030403 (2008) }
  \href{http://arxiv.org/abs/0801.1671}{\textcolor{magenta}{[arXiv:0801.1671 [hep-th]]}}.
   

\bibitem{Transmut} 
  A.~Arancibia and M.~S.~Plyushchay,
   \emph{``Transmutations of supersymmetry through soliton scattering, and self-consistent condensates,''}
    \href{http://journals.aps.org/prd/abstract/10.1103/PhysRevD.90.025008}{Phys.\ Rev.\ D {\bf 90}, no. 2, 025008 (2014)}
   \href{http://arxiv.org/abs/1401.6709}{\textcolor{magenta}{ [arXiv:1401.6709 [hep-th]]}}.

\bibitem{Dub}
S. Yu. Dubov, V. M. Eleonskii and N. E. Kulagin, 
 \emph{  ``Equidistant spectra of anharmonic oscillators,"}  
 \href{http://www.jetp.ac.ru/cgi-bin/e/index/r/102/3/p814?a=list}{
Zh. Eksp. Teor. Fiz. {\bf 102}, 814 (1992)};
\href{http://scitation.aip.org/content/aip/journal/chaos/4/1/10.1063/1.166056}{
Chaos {\bf 4}, 47 (1994)}.

 \bibitem{Adler}
 V. E. Adler, \emph{``A modification of Crum's method,"}
    \href{http://link.springer.com/article/10.1007%2FBF01035458}{
 Theor. Math. Phys. {\bf 101}, 1381 (1994)}.

\bibitem{SamOvc}
B. F. Samsonov and I. N. Ovcharov, 
``Darboux transformation and exactly solvable potentials with quasi-equidistant spectrum,"
\href{https://link.springer.com/article/10.1007/BF00559274}{Russ. Phys. J. {\bf 38}, 765 (1995)}.

\bibitem{BagSam}
V.ÊG.ÊBagrov and B.ÊF.ÊSamsonov, 
\emph{``Darboux transformation, factorization, and supersymmetry 
in one-dimensional quantum mechanics,"}  
\href{https://link.springer.com/article/10.1007%2FBF02065985}{
Theor. Math. Phys.  {\bf 104},  1051 (1995).}

\bibitem{Spir}
V. Spiridonov,
 {\it ``Universal superpositions of coherent states and self-similar potentials,"}
\href{http://journals.aps.org.ezproxy.usach.cl/pra/abstract/10.1103/PhysRevA.52.1909}{ Phys. 
Rev. A {\bf 52}, 1909 (1995)} 
\href{https://arxiv.org/abs/quant-ph/9601030}{\textcolor{magenta}{[arXiv:quant-ph/9601030]}}.

 \bibitem{JunRoy}
G. Junker  and  P. Roy, 
\emph{``Conditionally exactly solvable problems and non-linear algebras,"
\href{http://www.sciencedirect.com/science/article/pii/S0375960197004222}{Phys. 
Lett. A {\bf 232}, 155 (1997).}}

\bibitem{CPRS} 
J. F. Cari\~nena, A. M. Perelomov, M. F. Ra\~nada and M. Santander, 
 \emph{``A quantum exactly solvable nonlinear oscillator related
to the isotonic oscillator,"}
\href{http://iopscience.iop.org/article/10.1088/1751-8113/41/8/085301/meta}{
 J. Phys. A: Math. Theor. {\bf 41}, 085301 (2008) }
 \href{http://arxiv.org/abs/0711.4899}{\textcolor{magenta}{  [arXiv:0711.4899 [quant-ph]]}}.




\bibitem{FellSmi}
J. M. Fellows and R. A. Smith, 
 \emph{  ``Factorization solution of a family of quantum nonlinear oscillators,"}
 \href{http://iopscience.iop.org/article/10.1088/1751-8113/42/33/335303/meta}{
  J. Phys. A {\bf 42}, 335303 (2009) }.
  
  
\bibitem{Sesma}
J. Sesma, 
 \emph{  ``The generalized quantum isotonic oscillator,"} 
 \href{http://iopscience.iop.org/article/10.1088/1751-8113/43/18/185303/meta}{
J. Phys.  A
{\bf 43}, 185303} (2010)  
 \href{http://arxiv.org/abs/1005.1227}{\textcolor{magenta}{[arXiv:1005.1227 [quant-ph]]}}.
  
 
 \bibitem{GGM}
D. G\'omez-Ullate, Y. Grandati, and R. Milson, 
 \emph{``Rational extensions of
the quantum harmonic oscillator and exceptional Hermite polynomials,"}
\href{http://iopscience.iop.org/article/10.1088/1751-8113/47/1/015203/meta}{J.
 Phys. A {\bf 47}, 015203 (2014) }
 \href{http://arxiv.org/abs/1306.5143}{\textcolor{magenta}{[arXiv:1306.5143 [maph-ph]]}}. 

 
 \bibitem{Pupas}
 A. M. Pupasov-Maksimov,
 \emph{``Propagators of isochronous an-harmonic oscillators 
 and Mehler formula for the exceptional Hermite polynomials,"}
\href{http://www.sciencedirect.com.ezproxy.usach.cl/science/article/pii/S0003491615003565}{
Annals of Physics {\bf 363}, 122 (2015)}
\href{https://arxiv.org/abs/1502.01778}{\textcolor{magenta}{[arXiv:1502.01778 [maph-ph]]}}.  
  
\bibitem{Krein}
M. G. Krein,
\emph{ ``On a continuous analogue of a Christoffel 
 formula from the theory of orthogonal polynomials," }
 Dokl. Akad. Nauk SSSR {\bf 113}, 970 (1957).
 
 \bibitem{GomGrMil}
 D. Gomez-Ullate, Y. Grandati, and R. Milson,
 \emph{ ``Extended Krein-Adler theorem for the translationally shape 
 invariant potentials,"}
 \href{http://aip.scitation.org/doi/10.1063/1.4871443}{J. Math. Phys. {\bf 55}, 040201 (2014)}
 \href{http://arxiv.org/abs/1309.3756}{\textcolor{magenta}{[arXiv:1309.3756 [nlin.SI]]}}.

  
  \bibitem{Exc1}
  D. G\'omez-Ullate, N. Kamran, and  R. Milson,
 \emph{  ``An extended class of orthogonal polynomials defined by a Sturm-Liouville problem,"}
\href{http://www.sciencedirect.com/science/article/pii/S0022247X09004569}{
 J. Math. Anal. Appl. {\bf 359}
(2009) 352}
\href{https://arxiv.org/abs/0807.3939}{\textcolor{magenta}{[arXiv:0807.3939 [math-ph]]}}.

  \bibitem{Exc1+}
  D. G\'omez-Ullate, N. Kamran, and  R. Milson,
 \emph{    ``An extension of BochnerÕs problem: exceptional invariant subspaces,"} 
\href{http://www.sciencedirect.com/science/article/pii/S0021904509001853}{
J. Approx. Theory {\bf 162}, 897 (2010)}
\href{https://arxiv.org/abs/0805.3376}{\textcolor{magenta}{[arXiv:0805.3376 [math-ph]]}}.

\bibitem{OdSas}
 S. Odake and  R. Sasaki, 
 \emph{    ``Infinitely many shape invariant potentials and new 
 orthogonal polynomials, }
 \href{http://www.sciencedirect.com/science/article/pii/S0370269309009186}{
 Phys. Lett. B {\bf 679}, 414 (2009)}
 \href{https://arxiv.org/abs/0906.0142}{\textcolor{magenta}{[arXiv:0906.0142 [math-ph]]}}.
 
 \bibitem{OdSas+}
 S. Odake and R. Sasaki, 
 \emph{    ``Another set of infinitely many exceptional (X) 
 Laguerre polynomials, }
 \href{http://www.sciencedirect.com/science/article/pii/S0370269310000158}{
Phys. Lett. B {\bf 684}, 173 (2010)}
\href{https://arxiv.org/abs/0911.3442}{\textcolor{magenta}{[arXiv:0911.3442 [math-ph]]}}.

\bibitem{Ques+}
C. Quesne, 
 \emph{   ``Exceptional orthogonal polynomials, exactly 
 solvable potentials and supersymmetry,"} 
 \href{http://iopscience.iop.org/article/10.1088/1751-8113/41/39/392001/meta}{
J. Phys. A {\bf 41}, 392001 (2008) } 
\href{http://arxiv.org/abs/0807.4087}{\textcolor{magenta}{[arXiv:0807.4087 [quant-ph]]}}.

\bibitem{Ques++}
C. Quesne, 
 \emph{   ``Solvable rational potentials and exceptional 
orthogonal polynomials in supersymmetric quantum mechanics,"} 
\href{http://www.emis.de/journals/SIGMA/2009/084/}{
SIGMA {\bf 5}, 084  (2009)}
\href{https://arxiv.org/abs/0906.2331}{\textcolor{magenta}{[arXiv:0906.2331 [math-ph]]}}.


\bibitem{SasTsZh}
R. Sasaki, S. Tsujimoto and  A. Zhedanov, 
 \emph{    ``Exceptional Laguerre and Jacobi polynomials 
and the corresponding potentials through Darboux-Crum
 transformations,"}
 \href{http://iopscience.iop.org/article/10.1088/1751-8113/43/31/315204/meta}{
J. Phys. A {\bf 43}, 315204 (2010) }
\href{https://arxiv.org/abs/1004.4711}{\textcolor{magenta}{[arXiv:1004.4711 [math-ph]]}}.

\bibitem{Grand}
Y. Grandati, 
 \emph{   ``Solvable rational extensions of the isotonic oscillator,"}
 \href{http://www.sciencedirect.com/science/article/pii/S000349161100039X}{
Ann. Phys. {\bf  326}, 2074 (2011)  }
\href{https://arxiv.org/abs/1101.0055}{\textcolor{magenta}{[arXiv:1101.0055 [math-ph]]}}.

\bibitem{Sasaki} 
  R.~Sasaki,
  \href{http://www.apcospa.org/vol2/2-2.html}{The Universe {\bf 2}, no. 2, 2 (2014)}
  \href{https://arxiv.org/abs/1411.2703}{\textcolor{magenta}{[arXiv:1411.2703 [math-ph]]}}.


\bibitem{VesSha}
A. P. Veselov  and A. B. Shabat,   
{\it ``Dressing chains and the spectral theory of the Schr\"odinger operator,"} 
\href{http://link.springer.com/article/10.1007%2FBF01085979}{
Funct. Anal. Appl. {\bf 27}, 81 (1993)}. 

  
\bibitem{AdlPai}
 V. E. Adler,  {\it ``Nonlinear chains and Painlev\'e equations,"}
 \href{http://www.sciencedirect.com/science/article/pii/016727899490104X}{
 Physica D {\bf 73}, 335 (1994)}.
  
\bibitem{Hiet}
R. Willox and J. Hietarinta, {\it ``Painlev\'e equations from Darboux chains: I.  PIII - PV,"}
 \href{http://iopscience.iop.org/article/10.1088/0305-4470/36/42/014}{
J. Phys. A {\bf 36}, 10615 (2003)}.  

\bibitem{oblom}
A. Oblomkov, 
{\it ``Monodromy-free Schr\"odinger operators with quadratically
 increasing potentials,"} 
 \href{http://link.springer.com/article/10.1007/BF02557204}{Theor. Math. Phys. {\bf 121}, 1574 (1999)}.
 

\bibitem{MarQue1+}
I. Marquette and  C. Quesne,
  \emph{``Two-step rational extensions of the harmonic oscillator: exceptional orthogonal 
  polynomials and ladder operators," 
  \href{http://iopscience.iop.org/article/10.1088/1751-8113/46/15/155201/meta}{J. 
  Phys. A: Math. Theor. {\bf 46},  155201 (2013)}  
  \href{https://arxiv.org/abs/1212.3474}{\textcolor{magenta}{ [arXiv:1212.3474Ê[math-ph]]}}.}
  
  \bibitem{MarQue1}
  I. Marquette and  C. Quesne,
 \emph{``New ladder operators for a rational extension of the 
  harmonic oscillator and superintegrability of some two-dimensional systems,"}
\href{http://aip.scitation.org/doi/10.1063/1.4823771}{J. Math. Phys. {\bf 54}, 102102 (2013)}
\href{https://arxiv.org/abs/1303.7150}{\textcolor{magenta}{[arXiv:1303.7150 [math-ph]]}}.


\bibitem{MarQue1++}
 I.~Marquette and C.~Quesne,
  \emph{    ``Combined state-adding and state-deleting approaches
   to type III multi-step rationally-extended potentials: applications to ladder operators and 
   superintegrability,''}
  \href{http://aip.scitation.org/doi/10.1063/1.4901006}{J.\ Math.\ Phys.\  {\bf 55}, 112103 (2014)}
 \href{https://arxiv.org/abs/1402.6380}{\textcolor{magenta}{ [arXiv:1402.6380 [math-ph]]}}.


\bibitem{MarI}
I. Marquette,
  \emph{ ``New families of superintegrable systems from k-step rational 
extensions, polynomial algebras and degeneracies,"}
\href{http://iopscience.iop.org/article/10.1088/1742-6596/597/1/012057/meta}{J. Phys.: 
Conf. Ser. {\bf 597}, 012057 (2015)}
 \href{https://arxiv.org/abs/1412.0312}{\textcolor{magenta}{ [arXiv:1412.0312 [math-ph]]}}.

\bibitem{Ques}
C. Quesne,
 \emph{ ``Ladder operators for solvable potentials connected with 
 exceptional orthogonal polynomials,"}
 \href{http://iopscience.iop.org/article/10.1088/1742-6596/597/1/012064/meta}{J. Phys. 
 Conf. Ser. {\bf 597}, 012064 (2015)}
 \href{https://arxiv.org/abs/1412.5874}{\textcolor{magenta}{ [arXiv:1412.5874 [math-ph]]}}.

\bibitem{CarPly} 
  J.~F.~Cari\~nena and M.~S.~Plyushchay,
   \emph{``Ground-state isolation and discrete flows in a rationally extended quantum harmonic oscillator,"}
\href{http://journals.aps.org/prd/abstract/10.1103/PhysRevD.94.105022}{Phys.\ Rev.\ D 
{\bf 94}, no. 10, 105022 (2016)}
  \href{https://arxiv.org/abs/1611.08051}{\textcolor{magenta}{[arXiv:1611.08051 [hep-th]]}}.

\bibitem{SchwAn} 
  M.~S.~Plyushchay,
  \emph{``Schwarzian derivative treatment of the quantum second-order supersymmetry anomaly, and coupling-constant metamorphosis,''}
\href{http://www.sciencedirect.com/science/article/pii/S0003491616302743}{Ann. Phys., in press}  
 \href{https://arxiv.org/abs/1602.02179}{\textcolor{magenta}{[arXiv:1602.02179 [hep-th]]}}.


\bibitem{BurCha}
J.L. Burchnall, T.W. Chaundy,
{\it Commutative ordinary differential operators},
\href{http://onlinelibrary.wiley.com/doi/10.1112/plms/s2-21.1.420/abstract;jsessionid=0DB9784C385AC0EDE2AAD1190695C1AF.f02t01}{Proc. London Math. Soc. Ser. 2, {\bf 21}, 420 (1923)};
{\it Commutative ordinary differential operators},
\href{http://rspa.royalsocietypublishing.org/content/118/780/557}{Proc.  Royal Soc. London A {\bf 118}, 557  (1928)}.

\bibitem{Krich}
I. M. Krichever, 
  \emph{``Integration of nonlinear equations by the methods of 
  algebraic geometry,} 
  \href{http://link.springer.com/article/10.1007%2FBF01135528}{Func. Anal. Appl. {\bf 11}, 12 (1977)}; 
  \emph{``Baker-Akhiezer functions and integrable systems,"} 
  in:   \textsl{``Integrability. The Seiberg-Witten and Whitham Equations",} Edited by
  H. W. Braden and I. M. Krichever, p. 1 (Gordon and Breach Science Publishers, Amsterdam, 
  2000).
  

\end{thebibliography}
\end{document}